\documentclass{sig-alternate}
\usepackage{amsmath}
\usepackage{amsfonts,epsfig,subfigure,latexsym,graphicx}
\usepackage{multirow}
\usepackage{algorithm,algorithmic,bm}

\begin{document}

\title{Integrating Heterogeneous Information via Flexible Regularization Framework for Recommendation
}
%
%
%
%
%

\numberofauthors{3} 
%
\author{
%
%
\alignauthor
Chuan Shi\\
       \affaddr{Beijing University of Posts and
Telecommunications}\\
       \affaddr{Beijing, China}\\
       \email{shichuan@bupt.edu.cn}
\alignauthor
Jian Liu\\
       \affaddr{Beijing University of Posts and
Telecommunications}\\
       \affaddr{Beijing, China}\\
       \email{fullback@yeah.net}
\alignauthor
Fuzhen Zhuang\\
       \affaddr{Institute of Computing Technology, CAS}\\
       \affaddr{Beijing, China}\\
       \email{zhuangfz@ics.ict.ac.cn}
\and
\alignauthor
Philip S. Yu\\
       \affaddr{University of Illinois at Chicago}\\
         \affaddr{IL,USA}\\
       \email{psyu@cs.uic.edu}
\alignauthor
Bin Wu\\
       \affaddr{Beijing University of Posts and
Telecommunications}\\
       \affaddr{Beijing, China}\\
       \email{wubin@bupt.edu.cn}
}

\maketitle
\begin{abstract}
Recently, there is a surge of social recommendation, which leverages social relations among users to improve recommendation performance. However, in many applications, social relations are absent or very sparse. Meanwhile, the attribute information of users or items may be rich. It is a big challenge to exploit these attribute information for the improvement of recommendation performance.  In this paper, we organize objects and relations in recommendation system as a heterogeneous information network, and introduce meta path based similarity measure to evaluate the similarity of users or items. Furthermore, a matrix factorization based dual regularization framework SimMF is proposed to flexibly integrate different types of information through adopting the similarity of users and items as regularization on latent factors of users and items. Extensive experiments not only validate the effectiveness of SimMF but also reveal some interesting findings. We find that attribute information of users and items can significantly improve recommendation accuracy, and their contribution seems more important than that of social relations. The experiments also reveal that different regularization models have obviously different impact on users and items.
\end{abstract}



\keywords{Recommendation system, heterogeneous information network, matrix factorization,
similarity measure}

\section{Introduction}

In order to tackle information overload problem, recommender systems are proposed to help users to find objects of interest through utilizing the user-item interaction information and/or content information associated with users and items. Recommender systems have attracted much attention from multiple disciplines, and many techniques have been proposed to build recommender systems.  Among different recommendation techniques, hybrid recommendation \cite{AT05} is widely studied, which can achieve better recommendation performance in certain scenarios through combining user feedback data (e.g., ratings) and additional information of users or items. Particularly, with increasing popularity of social media, there is a surge of social recommendation techniques \cite{JE09,MZLK11}, which leverage rich social relations among users, such as friendships in Facebook, following relations in Twitter.

However, the emerging social recommendation usually faces the problem of relation sparsity. On the one hand, dense social relations can improve the recommendation performance. However, social relations are unavailable or very sparse in many real applications. For example, there are no social relations of uses in Amazon, and 80\% users in Yelp have less than 3 following relations.  On the other hand, users and items in many applications have rich attribute information, which  are seldom exploited. These information may be very useful to reveal users' taste and items' properties. For example, the group attribute of users can reflect their interests, and the type attribute of movies can reveal the content of movies. So it is desirable to effectively integrate all kinds of information for better recommendation performance, including not only feedback and social relations but also attributes of users and items. Some works have began to explore this issue \cite{JL13,YRSGSKNH14,YRSSKGNH13}, while they did not focus on revealing the importance of these attributes and their effect on recommendation accuracy.

\begin{figure}[htbp]
\centering
\small
\begin{minipage}[t]{0.27\textwidth}
  \includegraphics[width=4.5cm]{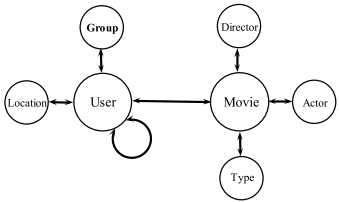}
\end{minipage}%
\vspace{-10pt}
\caption{\small Objects and relations in movie recommendation system is organized as a heterogeneous information network. }\label{fig:fft}
\end{figure}

Although integrating more information is promising to achieve better recommendation performance, how to integrate these information still faces two challenges. (1) The information to be integrated has different types. These mixed information types include integer number (i.e., rating information), vector (i.e., attribute information), and graph (i.e., social relations). We need to design a unified model to effectively integrate these different types of information. (2) A unified and flexible method is desirable to integrate all or some of these information. In order to intensively study the impact of different information, designed method should flexibly integrate different granularity of information and uniformly utilize different types of information.

In this paper, we organize the objects and relations in recommendation system as a heterogeneous information network which contains different types of nodes or links.  Figure 1 shows such an example representing the objects and their relations in a movie recommendation system (detailed in Section 3). Intuitively, network can effectively integrate different types of heterogeneous information including not only feedback (i.e., user-movie) and social relations (i.e., user-user) but also attribute information of users (e.g., user-group) and items (e.g., movie-type). Moreover, meta path, a  relation composition connecting two types of objects, contains rich semantics information \cite{SHYYW11}. For example, the meta path ``User-Movie-User'' connecting users means users viewing the same movies.

In order to utilize these heterogeneous information, we introduce meta path based similarity measure to evaluate the similarity of users and items. Based on matrix factorization, a dual regularization framework SimMF is proposed to integrate heterogeneous information through adopting similarity information of users and items as regularization on latent factors of users and items. Moreover, in SimMF, two different regularization models, average and individual based regularization, can flexibly confine regularization on users or items. Extensive experiments on four real datasets (i.e., Douban Movie, Yelp, MovieLens and Douban Book) validate the effectiveness of SimMF and reveal some interesting and useful findings. The major contributions of this paper are summarized as follows:

(1) We proposed a unified and flexible matrix factorization based dual regularization framework to integrate heterogeneous information. The framework can flexibly and granularly integrate different types of information. In addition, it provides two optional regularization models on users and items.

(2) We crawled comprehensive Douban Movie and Douban Book datasets including feedback, social relations and attribute information of users and items. More importantly, extensive experiments reveal some interesting and useful findings. On these experimental datasets, the attribute information of items and users can significantly enhance recommendation performance. Their improvements are even higher than that of social relations. In addition, the similarity information generated by meta paths with dense relations and meaningful semantics usually obtain better performance. These findings indicate that, although social recommendation is an important direction, how to effectively utilize attribution information may also be a promising way to further improve recommendation performance.

(3) Another important finding is that different regularization models on users and items have obvious effects on recommendation performance. Ma et al. \cite{MZLLK11} have studied the effect of different regularization models on social relations, we further discuss the effect on similarity relations of users and items. This finding illustrates that it is helpful to set proper regularization model according to data property in real applications.

The remainder of this paper is organized as follows. Section 2 introduces the related work. Section 3 presents some preliminary knowledge, then the proposed SimMF model is detailed in Section 4. Experiments and analysis are shown in Section 5. Last, we conclude the paper in Section 6.

\section{Related work}
According to the utilized information for recommendation, we can roughly classify contemporary recommendation methods into three types: feedback based, social relation based and heterogeneous information based methods.

Traditional recommender systems normally only utilize user-item rating feedback information for recommendation. Collaborative filtering is one of the most popular techniques, which includes two types of approaches: memory-based method and model-based method. Recently, matrix factorization has shown its effectiveness and efficiency in recommender systems, which factorizes user-item rating matrix into two low rank user-specific and item-specific matrices, then utilizes the factorized matrices to make further predictions \cite{SJ03}.

With the prevalence of social media, more and more research study social recommender systems which utilize social relations among users. Many researchers utilized trust information among users. Ma et al. \cite{MYLK08} fused user-item matrix with users' social trust networks by sharing a common latent low-dimensional user feature matrix. Furthermore, the authors in \cite{MKL11} coined the social trust ensemble to represent the formulation of the social trust restrictions. More and more researches began to use friend relation among users. In \cite{MZLLK11}, the additional social regularization term ensures that the distance of latent feature vectors of two friends with similar tastes to be closer. Yang et al. \cite{YSL12} inferred category-specific social trust circles from available rating data combined with friend relations. Recently, many studies have begun to utilize other types of informations. For example, Cantador et al. \cite{IAD10} made use of user and item profiles defined in terms of weighted lists of social tags for top N recommendation. Furthermore, they presented a comparative study on the influence that different types of information available in social systems have on item recommendation \cite{RIP13}.

Research on heterogeneous information network, in which objects are of different types and links among objects represent different relations, has surged over the years. More and more researchers have been aware of the importance of heterogeneous information for recommendation. Jones et al. \cite{CJA11} validated the importance of the exploitation on available heterogeneous data sources and proposed a Bayesian approach called LaD-BAE to capture both feature heterogeneity and predictive heterogeneity. Zhang et al. \cite{ZTL08} investigated the problem of recommendation over heterogeneous network and formalized the recommendation as a ranking problem then proposed a random walk model to estimate the importance of each object in the heterogeneous network. Considering heterogeneous network constructed by different interactions of users, Jamali and Lakshmanan \cite{JL13} proposed HETEROMF to integrate a general latent factor and context-dependent latent factors. Wang et al. \cite{FW12} proposed the OptRank method to alleviate the cold start problem by utilizing heterogeneous information contained in social tagging system. Yu et al. \cite{YRSGSKNH14,YRSSKGNH13} proposed an implicit feedback recommendation model with systematically extracted latent features from heterogeneous network. Furthermore, they utilized users' clicked URLs to build a Freebase entity graph, which is a heterogeneous information network \cite{YMHH14}. More recently, Luo et al. \cite{LPW14} proposed a collaborative filtering-based social recommendation method, called Hete-CF, using heterogeneous relations, and Burke et al. \cite{RFB14}  incorporated multiple relations generated by meta paths in a weighted hybrid model. Vahedian \cite{FV14} designed the WHyLDR approach for multiple recommendation tasks, which combines heterogeneous information with a linear-weighted hybrid model. In addition, due to massive amounts of fashion items available online, Hanbit et al. \cite{LHLS15} extracted meta-paths from heterogeneous information network and designed a meta-path based method for fashion items recommendation. Shi et al. \cite{SZLYYW15} proposed the concept of weighted heterogeneous information network and designed a meta-path based recommendation model called SemRec.

The proposed SimMF belongs to heterogeneous information based methods. Compared to feedback based and social relation based methods, SimMF can flexibly integrate various heterogeneous information. And SimMF is also different from existing heterogeneous information based models in several aspects. Contemporary methods usually consider one or two types of heterogeneous information. For example, HETEROMF focuses on different interactions of users. The method proposed by Yu et al. only considers attributes of items \cite{YRGSH13}, and it is an item recommendation \cite{YRSGSKNH14,YRSSKGNH13} model. SimMF considers all kinds of information and flexibly integrates them together. Moreover, we intensively investigate the impact of this heterogeneous information which is seldom explored before. WHyLDR considers heterogeneous information as SimMF does. However, while WHyLDR focuses on component selection and component combination and is for item recommendation rather than rating prediction. The method proposed in \cite{LHLS15} and SemRec \cite{SZLYYW15} are both meta-path based model, while SimMF should be considered as a matrix factorization based model. The proposed work is similar to Hete-CF, but Hete-CF only applies one type of matrix factorization constraint called individual regularization on users and items, SimMF considers two types of regularization and exploits their different impacts on recommendation performance.

\section{Preliminary}
In this section, we describe notations used in this paper and present some preliminary knowledge.

A Heterogeneous Information Network (HIN) is a special type of information network with underneath data structure as a directed graph, which either contains multiple types of objects or multiple types of links. Specifically, given a schema $S=(\mathcal{A},\mathcal{R})$ which consists of a set of entity types $\mathcal{A} = \{A\}$ and a set of relations $\mathcal{R} = \{R\}$, an information network is defined as a directed graph $G = (V,E)$ with an object type mapping function $\varphi : V \rightarrow \mathcal{A}$ and a link type mapping function $\psi : E \rightarrow \mathcal{R}$. If types of objects $|\mathcal{A}| > 1$ or types of relations $|\mathcal{R}| > 1$, the network is called \textbf{heterogeneous information network}; otherwise, it is a \textbf{homogeneous information network}.

Figure 1 shows the network schema of a typical heterogeneous network which organizes objects and relations in movie recommendation system. The heterogeneous network contains objects from multiple types of entities:
user (U), movie (M), group (G), location (L), actor (A), director (D), and type (T).
For each user, it has links to a set of
other users as his (her) friends, a set of affiliated groups, and a set of rated movies. Links exist between user and user denoting the friendship relation, between user and group denoting
the membership relation, between user and movie denoting rating and rated relation. It is similar for movie. We can find that this HIN includes different types of information, such as feedback (i.e., user-movie), social relations (i.e., user-user), and attributes (e.g., user-group, movie-actor).

Two objects in a heterogeneous network can be connected via different paths, which can be called \textbf{meta path} \cite{SHYYW11}. A meta path $\mathcal{P}$ is a path defined on a schema $S=(\mathcal{A},\mathcal{R})$, and is denoted in the form of $A_1 \xrightarrow{R_1} A_2 \xrightarrow{R_2} \cdots \xrightarrow{R_l} A_{l+1}$ (abbreviated as $A_1A_2\cdots A_{l+1}$), which defines a composite relation $R = R_1 \circ R_2 \circ \cdots \circ R_l$ between type $A_1$ and $A_{l+1}$, where $\circ$ denotes the composition operator on relations. As an example shown in Figure 1, users can be connected via ``User-User'' (UU) path, ``User-Group-User'' (UGU) path, ``User-Movie-User'' (UMU) and so on. It is obvious that semantics underneath these paths are different. The UU path means users' friends (i.e., friend relation among users), while the UMU path means users watching the same movies. Since different meta paths have different semantics, objects connecting by different meta paths have different similarity. So we can evaluate the similarity of users (or movies) based on different meta paths. For example, for users,we can consider meta paths UU, UGU, UMU and so on. Similarly, meaningful meta paths connecting movies include MAM, MDM, and so on.

There are several path based similarity measures to evaluate the similarity of objects in HIN \cite{LC10a,SKHYW14,SHYYW11}. Considering semantics in meta paths, Sun et al. \cite{SHYYW11} proposed PathSim to measure the similarity of same-typed objects
based on symmetric paths. Lao and Cohen \cite{LC10a} proposed a Path Constrained Random Walk (PCRW) model to measure the entity proximity in a labeled directed graph constructed by the rich
metadata of scientific literature. The HeteSim \cite{SKHYW14} can measure the relatedness of heterogeneous objects based on an arbitrary meta path. All these similarity measures can be used in the similarity calculation, and their differences can be seen in reference \cite{SKHYW14}.

We define $S_{ij}^{(l)}$ to denote the similarity of two objects $u_i$ and $u_j$ under the given meta path $\mathcal{P}_l$. The similarity ($\mathcal{S}$) is determined by the given meta path ($\mathcal{P}$) and the similarity measure ($\mathcal{M}$). That is $\mathcal{S}=\mathcal{P}\times \mathcal{M}$. We know that the similarity of different paths are different and they are incomparable. So we normalize them with \emph{Sigmoid} function as shown in Equation 1, where $\bar{S}^{(l)}$ means the average of $S_{ij}^{(l)}$ and $\beta$ is set to 1. The normalization process has the following two advantages. (1) It confines the similarity into $[0,1]$ without changing their ranking. (2) It can reduce the similarity difference of different paths. In the following section, we directly use the $S_{ij}^{(l)}$ to represent the normalized similarity.

\vspace{-10pt}
\begin{equation}\label{}\scriptsize
{S_{ij}^{(l)}}^{'}=\frac{1}{1+e^{-\beta\times (S_{ij}^{(l)}-\bar{S}^{(l)})}}
\end{equation}

Since users (or items) have different similarity under different meta paths, we consider their similarity on all paths through assigning weights on different paths. For users, we define $S^\mathcal{U}$ for the similarity matrix of users on all paths, and $S^\mathcal{I}$ for the similarity matrix of items on all paths. They can be defined as follows, where $w^\mathcal{U}_l$ represents the weight of similarity matrix of users under the path $\mathcal{P}_l$ and $w^\mathcal{I}_l$ represents that of items.

\vspace{-10pt}
\begin{equation}\label{}\scriptsize
\begin{array}{ll}
S^\mathcal{U}=\sum_{l}{w^\mathcal{U}_lS^{(l)}}  &\Sigma_{l}{w^\mathcal{U}_l}=1; 0\leq w^\mathcal{U}_l\leq 1  \\
S^\mathcal{I}=\sum_{l}{w^\mathcal{I}_lS^{(l)}}  &\Sigma_{l}{w^\mathcal{I}_l}=1; 0\leq w^\mathcal{I}_l\leq 1
\end{array}
\end{equation}
\vspace{-10pt}


\section{The SimMF Method}
In this section, we will introduce the \textbf{SimMF} method which utilizes \textbf{m}atrix \textbf{f}actorization framework to incorporate \textbf{sim}ilarity information. We firstly review the basic low-rank matrix factorization framework, and then introduce the improved model through constraining similarity regularization on users and items, respectively. Finally, we show the unified model through applying similarity regularization on users and items simultaneously.

\subsection{Low-Rank Matrix Factorization}
The low-rank matrix factorization has been widely studied in recommendation system \cite{SJ03}.
Its basic idea is to factorize the user-item rating matrix $R$ into two matrices ($U$ and $V$) representing  users' and items' distributions on latent semantic, respectively. Then, the rating prediction can be made through these two specific matrices. Assuming an $m \times n$ rating matrix $R$ to be $m$ users' ratings on $n$ items, this approach mainly minimizes the objective function $\mathcal{L}(R,U,V)$ as follows.

\vspace{-10pt}
\begin{eqnarray}\scriptsize
\min_{U,V}\mathcal{L}(R,U,V)&=&\frac{1}{2}\sum_{i=1}^{m}\sum_{j=1}^{n}I_{ij}(R_{ij} - U_i V^T_j)^2\nonumber \\
                            & &+\frac{\lambda_1}{2}\|U\|^2+\frac{\lambda_2}{2}\|V\|^2
\end{eqnarray}
where $I_{ij}$ is the indicator function that is equal to 1 if user $i$ rated item $j$ and equal to 0 otherwise. $U\in \mathbb{R}^{m \times d}$ and $V\in \mathbb{R}^{n \times d}$, where $d$ is the dimension number of latent factors and $d \ll min(m,n)$.
$U_i$ is a row vector derived from $i$th row of matrix $U$ and $V_j$ is a row vector derived from $j$th row of matrix $V$. $\lambda_1$ and $\lambda_2$ represent the regularization parameters. In summary, the optimization problem minimizes the sum-of-squared-errors objective function with quadratic regularization terms which aim to avoid overfitting. This problem can be effectively solved by a simple stochastic gradient descent technique.

\subsection{Similarity regularization on users}
As mentioned above, the user-specific factorized matrix describes users' distribution on latent semantic. In this section, we will introduce two different types of similarity regularization (i.e., average-based and individual-based regularization) on users to force distance between $U_p$ and $U_q$ to be much smaller if user $p$ is highly similar to user $q$.

\subsubsection{Average-based Regularization}
Intuitively, we have similar behavior model with people who are similar with us. That is, the latent factor of a user is similar to the latent factor of people who are the most similar to the user. Based on this assumption, we add user's similarity regularization to the basic low-rank matrix factorization framework.

\vspace{-10pt}
\begin{eqnarray}\scriptsize
\min_{U,V}\mathcal{L}(R,U,V)&=&\frac{1}{2}\sum_{i=1}^{m}\sum_{j=1}^{n}I_{ij}(R_{ij} - U_i V^T_j)^2\nonumber \\
            & &+\frac{\alpha}{2}\sum_{i=1}^{m}\|U_i-\frac{\sum_{f\in \mathcal{T}_{u}^+(i)}S^\mathcal{U}_{if}U_f}{\sum_{f\in \mathcal{T}_{u}^+(i)}S^\mathcal{U}_{if}} \|^2
            \nonumber \\
            & &+\frac{\lambda_1}{2}\|U\|^2+\frac{\lambda_2}{2}\|V\|^2
\end{eqnarray}
where $\mathcal{T}_{u}^+(i)$ is the set of users who are in the top $k$ similarity list of user $i$ and $S^\mathcal{U}_{if}$ is the element located on $i$th row and $f$th column of user similarity matrix $S^\mathcal{U}$.  The average-based regularization confines that the latent factor of a user is close to the average of the latent factor of the top $k$ similar people to the user. The similar regularization has been used in social recommendation \cite{MZLLK11}, while it just limits to the friend list of users. Here the average-based regularization not only extends to the top $k$ similarity list of users but also considers the similarity values as the weights. The parameter $k$ can be set to tradeoff accuracy and computation cost. Large $k$ usually means high accuracy but low efficiency. In our experiments, $k$ is set to 5\% of the vector dimension. A local minimum of the objective function given by Equation 4 can be solved by performing gradient descent in feature vectors $U_i$ and $V_j$, which is shown in Equation 5 and 6. Here $\mathcal{T}_{u}^-(i)$ represents the set of users whose top $k$ similarity list contains user $i$.
\begin{eqnarray}\scriptsize
\frac{\partial \mathcal{L}}{\partial U_i}&=&\sum_{j=1}^{n}I_{ij}(U_iV_j^T-R_{ij})V_j
                                        \nonumber \\
                                        & &+\alpha(U_i-\frac{\sum_{f\in \mathcal{T}_{u}^+(i)}(S^\mathcal{U}_{if}U_f)}{\sum_{f\in \mathcal{T}_{u}^+(i)}S^\mathcal{U}_{if}})
                                        \\
                                        & &+\alpha\sum_{g\in \mathcal{T}_{u}^-(i)}\frac{-S^\mathcal{U}_{ig}(U_g-\frac{\sum_{f\in \mathcal{T}_{u}^+(g)}(S^\mathcal{U}_{gf}U_f)}
                                        {\sum_{f\in \mathcal{T}_{u}^+(g)}S^\mathcal{U}_{gf}})}{\sum_{f\in \mathcal{T}_{u}^+(g)}S^\mathcal{U}_{gf}}
                                        +\lambda_1U_i \nonumber
\end{eqnarray}
\begin{eqnarray}\scriptsize
\frac{\partial \mathcal{L}}{\partial V_j}&=&\sum_{i=1}^{m}I_{ij}(U_i V_j^T-R_{ij})U_i+\lambda_2V_j
\end{eqnarray}
\subsubsection{Individual-based Regularization}
The above average-based regularization constrains user's taste with the average
taste of people who are the most similar users. However, it may be ineffective for users whose similar users have diverse tastes. In order to avoid this disadvantage, we employ individual-based regularization on users as follows.
\vspace{-10pt}
\begin{eqnarray}\scriptsize
\min_{U,V}\mathcal{L}(R,U,V)&=&\frac{1}{2}\sum_{i=1}^{m}\sum_{j=1}^{n}I_{ij}(R_{ij} - U_i V^T_j)^2
                            \nonumber \\
                             & &+\frac{\alpha}{2}\sum_{i=1}^{m}\sum_{j=1}^{m}S^\mathcal{U}_{ij}\|U_i-U_j\|^2
                             \nonumber \\
                             & &+\frac{\lambda_1}{2}\|U\|^2+\frac{\lambda_2}{2}\|V\|^2
\end{eqnarray}
In essential, the individual-based regularization enforces a large $S^\mathcal{U}_{ij}$ to have a small distance between $U_i$ and $U_j$. That is, similar users have smaller distance on latent factors. With the same optimization technique, a local minimum of Equation 7 can also be found by performing gradient descent in $U_i$ and $V_j$.
\begin{eqnarray}\scriptsize
\frac{\partial \mathcal{L}}{\partial U_i}&=&\sum_{j=1}^{n}I_{ij}(U_i V_j^T-R_{ij})V_j\nonumber \\
   & &+\alpha\sum_{j=1}^{m}(S^\mathcal{U}_{ij}+S^\mathcal{U}_{ji})(U_i-U_j)+\lambda_1U_i
\end{eqnarray}
\vspace{-20pt}
\begin{eqnarray}
\frac{\partial \mathcal{L}}{\partial V_j}=\sum_{i=1}^{m}I_{ij}(U_i V_j^T-R_{ij})U_i+\lambda_2V_j
\end{eqnarray}


\begin{table*}[htbp]\scriptsize
\centering
\caption{\label{comparison}Statistics of Datasets}
\begin{tabular}{|c|c|c|c|c|c|c|c|}
  \hline
 {Datasets} & {Relation} & {Relations} &{number} & {number} & {number} & {min./max./ave.} & {min./max./ave.} \\
  & {type} & {(A-B)} &{of A} & {of B} & {of Relations} & { degrees of A} & { degrees of B} \\
  \hline
  \hline
  \multirow{7}{*}{Douban Movie}
  & {rating} & {User-Movie} & {13616} & {34453} & {1301072} & {1/818/95.6} & {1/3697/37.8} \\
  \cline{2-8}
  & {social relation}& {User-User} &{3198} & {3198} & {3129} & {1/49/2.0} & {1/49/2.0} \\
  \cline{2-8}
  & {attribute} &{User-Group} & {13582} & {2796} & {579555} & {1/499/42.7} & {50/12892/207.3} \\
  \cline{3-8}
  & {of users} & {User-Location}& {11463} & {354} & {11463} & {1/1/1} & {2/1690/32.4} \\
  \cline{2-8}
  &{attribute}&{Movie-Director} & {7916} & {964} & {8654} & {1/32/1.1} & {5/62/9.0} \\
  \cline{3-8}
  &{of}&{Movie-Actor} & {15488} & {3330} & {37539} & {1/4/2.4} & {4/101/11.3} \\
  \cline{3-8}
  & {movies} &{Movie-Type} & {29250} & {45} & {59990} & {1/3/2.1} & {1/14303/1333.1} \\
  \hline
  \hline
  \multirow{4}{*}{Yelp}
  &{rating} & {User-Business} & {14085} & {14037} & {194255} & {1/639/4.6} & {1/1026/20.7} \\
  \cline{2-8}
  &{social relation}& {User-User} &{9581} & {9581} & {150532} & {1/2032/10.0} & {1/2032/10.0} \\
  \cline{2-8}
  &{attribute}&{Business-Category} & {14037} & {575} & {39406} & {1/10/2.8} & {1/5556/73.9} \\
  \cline{3-8}
  &{of business}& {Business-Location}& {14037} & {62} & {14037} & {1/1/1.0} & {1/5493/236.1} \\
  \hline
  \hline
  \multirow{5}{*}{MovieLens}
 &{rating} & {User-Movie} & {6040} & {3952} & {180037} & {1/394/29.8} & {1/640/52.0} \\
  \cline{2-8}
  &{attribute}& {User-Gender} &{6040} & {2} & {6040} & {1/1/1.0} & {1709/4331/3020.0} \\
  \cline{3-8}
  &{of}&{User-Age} & {6040} & {7} & {6040} & {1/1/1.0} & {222/2096/862.8} \\
  \cline{3-8}
  &{users}& {User-Occupation} & {6040} & {21} & {6040} & {1/1/1.0} & {17/759/287.6} \\
  \cline{2-8}
  &{attribute of movies}& {Movie-Type} & {3952} & {18} & {6408} & {1/6/1.6} & {44/1603/356.0} \\
  \hline
  \hline
  \multirow{7}{*}{Douban Book}
 &{rating} & {User-Book} & {13024} & {22347} & {792026} & {1/2551/60.81} & {6/2679/35.4} \\
  \cline{2-8}
  &{social relation}& {User-User} &{12748} & {12748} & {169150} & {1/1998/13.3} & {1/1998/13.3} \\
  \cline{2-8}
  &{attribute}&{User-Group} & {13024} & {2936} & {1189271} & {1/629/91.3} & {100/13022/405.1} \\
  \cline{3-8}
  &{of users}& {User-Location} & {10592} & {453} & {10592} & {1/1/1.0} & {1/2480/23.4} \\
  \cline{2-8}
  &{attribute}& {Book-Author} & {21907} & {10805} & {21907} & {1/1/1.0} & {1/199/2.0} \\
  \cline{3-8}
  &{of}& {Book-Publisher} & {21773} & {1815} & {21773} & {1/1/1.0} & {1/981/11.9} \\
  \cline{3-8}
  &{books}& {Book-Year} & {21192} & {64} & {21192} & {1/1/1.0} & {1/2039/331.1} \\
  \hline
\end{tabular}
\end{table*}

\subsection{Similarity regularization on items}
For simplicity, we define the $Reg^x_y$ notation to represent the average-based or individual-based regularization term on users or items,  where $x\in\{\mathcal{U},\mathcal{I}\}$ means $\mathcal{U}$sers or $\mathcal{I}$tems and $y\in\{ave,ind\}$ means \emph{ave}rage or \emph{ind}ividual-based regularization. That is, for similarity regularization on users, we have
\vspace{-5pt}
\begin{eqnarray}\scriptsize
Reg^{\mathcal{U}}_{ave}&=&\sum_{i=1}^{m}\|U_i-\frac{\sum_{f\in \mathcal{T}_{u}^+(i)}S^\mathcal{U}_{if}U_f}{\sum_{f\in \mathcal{T}_{u}^+(i)}S^\mathcal{U}_{if}} \|^2\\
Reg^{\mathcal{U}}_{ind}&=&\sum_{i=1}^{m}\sum_{j=1}^{m}S^\mathcal{U}_{ij}\|U_i-U_j\|^2
\end{eqnarray}

Similar to the similarity regularization on users, we can also define these two different types of regularization on items as follows.
\vspace{-10pt}
\begin{eqnarray}\scriptsize
Reg^{\mathcal{I}}_{ave}&=&\sum_{j=1}^{n}\|V_j-\frac{\sum_{f\in \mathcal{T}_{i}^+(j)}S^\mathcal{I}_{jf}V_f}{\sum_{f\in \mathcal{T}_{i}^+(j)}S^\mathcal{I}_{jf}} \|^2 \\
Reg^{\mathcal{I}}_{ind}&=&\sum_{i=1}^{n}\sum_{j=1}^{n}S^\mathcal{I}_{ij}\|V_i-V_j\|^2
\end{eqnarray}
where $\mathcal{T}_{i}^+(j)$ is the set of items who are in the top $k$ similarity list of item $j$, and $S^\mathcal{I}_{jf}$ is the element located on $j$th row and $f$th column of similarity matrix $S^\mathcal{I}$. We can also define the optimization function based on these two regularization terms on items and derive their gradient learning algorithms as above.

\subsection{A Unified Dual Regularization}
Now we consider regularization on users and items simultaneously. The corresponding optimization function is shown as follows.
\begin{eqnarray}\scriptsize
\min_{U,V}\mathcal{L}(R,U,V)&=&\frac{1}{2}\sum_{i=1}^{m}\sum_{j=1}^{n}I_{ij}(R_{ij} - U_i V^T_j)^2
                        \nonumber \\
                        & &+\frac{\alpha}{2}Reg^{\mathcal{U}}_y+\frac{\beta}{2}Reg^{\mathcal{I}}_y \nonumber \\
                       & &+\frac{\lambda_1}{2}\|U\|^2+\frac{\lambda_2}{2}\|V\|^2
\end{eqnarray}
where $\alpha$ and $\beta$ control the ratio of user and item regularization, respectively. For $y\in\{ave,ind\}$, there are four regularization models. Similarly, we can use the gradient descent method to solve this optimization problem. The whole algorithm framework is shown in Algorithm 1.

\begin{algorithm}[htbp]
    \algsetup{linenosize=\small}
    \caption{Algorithm Framework of SimMF}
    \begin{algorithmic}[1]
    \small
    \REQUIRE ~~\\
        $G$: heterogeneous information network\\
        $\mathcal{P}_U,\mathcal{P}_I$: sets of meta paths related to users and items\\
        $\eta$: learning rate for gradient descent\\
        $\alpha,\beta,\lambda_1,\lambda_2$: controling parameters defined above\\
        $\epsilon$: convergence tolerance
    \ENSURE ~~\\
        $U,V$: the latent factor of users and items\\
    ~\\
    \STATE Calculate similarity matrix of user $S_\mathcal{U}$ based on $\mathcal{P}_U$, $G$
    \STATE Calculate similarity matrix of item $S_\mathcal{I}$ based on $\mathcal{P}_I$, $G$
    \STATE Initialize $U,V$
    \REPEAT
        \STATE $U_{old}:=U$,$V_{old}:=V$
        \STATE Calculate $\frac{\partial \mathcal{L}}{\partial U}$, $\frac{\partial \mathcal{L}}{\partial V}$
        \STATE Update $U:=U-\eta*\frac{\partial \mathcal{L}}{\partial U}$
        \STATE Update $V:=V-\eta*\frac{\partial \mathcal{L}}{\partial V}$
    \UNTIL {$\|U-U_{old}\|^2+\|V-V_{old}\|^2<\epsilon$}
    \end{algorithmic}
\end{algorithm}

\subsection{Discussion}
Through employing dual regularization on users and items,  SimMF is a general and flexible framework for matrix factorization based recommendation, which can integrate rating, social relations and attribute information of users and items. The $\alpha$ and $\beta$ control how much SimMF integrates information from social relations and attribute of users and items, and $S^\mathcal{U}$ and $S^\mathcal{I}$ decide what kind of similarity information will be used. If the $\alpha$ and $\beta$ both are set with 0, SimMF degrades to traditional collaborative filtering with matrix factorization \cite{SJ03}. When the $\alpha$ is 0, SimMF can integrate the attributes of items, which is recently considered by Yu et al. \cite{YRSGSKNH14,YRSSKGNH13}. When $\beta$ is 0, SimMF can fuse the social relations, like social recommendation \cite{MYLK08}, as well as the attributes information of users. Particularly, social relations in social recommendations can be presented through setting the similarity matrix of users $S^\mathcal{U}$ with the similarity generated by special meta paths. For example, in Douban Movie dataset, the friend relation can be represented by the meta path UU, and the membership can be represented by the meta path UGU. In this condition, SimMF converts to the social recommendation \cite{MKL11, MYLK08} indeed. In addition, SimMF considers two regularization models (i.e., individual and average based regularization) to integrate similarity information. We can find that these two regularization models have different impacts on users and items in the following experiments.

Let's give more discussion on the similarity matrix of users ($S^\mathcal{U}$) and items ($S^\mathcal{I}$). As we know,  $S^\mathcal{U}$ and $S^\mathcal{I}$ are the similarity matrix of users and items on multiple meta paths, respectively. There are two notable problems.  (1) How to select the meta paths  for users or items?  We know that there are infinite meta paths connecting users or items. As illustrated in the following experiments, the short and meaningful meta paths are helpful to achieve better recommendation performance through generating good similarity measures. Sun et al. \cite{SHYYW11}  pointed out that the semantics of long meta paths are not meaningful and they fail to produce good similarity measures.  Some priori knowledge  can also be applied to the selection of meta paths, such as domain knowledge and user-guided information \cite{Sun4}. (2) How to combine the multiple meta paths?  We can set proper weights for meta paths according to their importances.  Supervised weight learning can also be designed to automatically determine the weight of meta paths, as Yu et al. \cite{YRSGSKNH14} and Lao et al. \cite{LC10a} did. In this paper, we simply set the weight with the equal value, since the mean weight is sufficient to show the benefits of SimMF.

According to Algorithm 1, the complexity of SimMF can be analyzed as follows. SimMF contains two main parts: (1) Similarity evaluation (Lines 1-2). It can be completed offline and many strategies \cite{SKHYW14} can speed it up. (2) Parameters learning (Lines 4-9). The main computation of the parameters learning is to calculate the gradients. The complexity of calculating gradients need to consider two conditions: average-based and individual-based regularizations. Assume that $|R|$ is the number of nonzero entries in rating matrix $R$.  In terms of user related gradient, $|\mathcal{T}_{u}^-(i)|$ and $|\mathcal{T}_{i}^-(j)|$ can be usually estimated by a small constant $c$ and $c \ll m, c \ll n$. Thus, the complexity for average-based regularization $\frac{\partial \mathcal{L}}{\partial U}$ is $O((m \times k \times c+|R|) \times d)$ and the complexity for individual-based regularization  $\frac{\partial \mathcal{L}}{\partial U}$ is $O((m \times k+|R|) \times d)$. Similarly, the complexity for average-based regularization  $\frac{\partial \mathcal{L}}{\partial V}$ is $O((n \times k \times c+|R|) \times d)$ and the complexity for individual-based regularization $\frac{\partial \mathcal{L}}{\partial V}$ is $O((n \times k+|R|) \times d)$. In summary, the whole complexity of parameters learning is $O(((m+n) \times k \times c+|R|) \times d \times t)$ where $t$ is the number of iterations.

\section{Experiments}

In this section, we will verify the superiority of our model by conducting a series of experiments compared to state-of-the-art recommendation methods.

\subsection{Datasets}

Although there are many public datasets for recommendation, they focus on the rating information and social relations \cite{MYLK08, MZLLK11,MZLK11}. Yu et al. \cite{YRSGSKNH14,YRSSKGNH13} considered the attribute information of items, while they ignore the attribute information of users. In order to get more comprehensive heterogeneous information, including rating information, attribute information of users and items and social relations, we prepared four different datasets from three various domains.
\par Douban Movie \footnote{http://movie.douban.com/} and MovieLens \footnote{http://grouplens.org/datasets/movielens/} \cite{YRGSH13} are from the movie domain. Douban is a well known social media network in China. Douban Movie dataset includes 13,616 users and 34,453 movies with 1,301,072 movie ratings ranging from 1 to 5. Moreover, we also extract social realtions among users and attribute information of users (e.g., groups and locations) and movies (e.g., actors, directors and types). The network schema of Douban Movie is shown in Figure 1. MovieLens dataset contains rating information of users on movies and attributes information of user (e.g. age range and occupations). Stemming from the business domain, the widely-used Yelp challenge dataset \footnote{http://www.yelp.com/dataset\_challenge/} \cite{YRSGSKNH14,YRSSKGNH13} records users' ratings on local business and also contains social relations and attribute information of business (e.g. cities and categories). Belonging to the book domain,  the Douban Book \footnote{http://book.douban.com/} includes 13,024 users, 22,347 books, and 792,026 rating records between users and books. The detailed description can be seen in Table 1. Besides different domains, we can find that these four datasets have different characteristics. MovieLens dataset has dense rating information but with no social relation, and Douban Movie dataset has medium dense rating information with sparse social relations. In addition, Douban Book dataset has medium dense rating information with dense social relations, and Yelp dataset has sparse rating information with dense social relations.

\subsection{Metrics}
We use Mean Absolute Error (MAE) and Root Mean Square Error (RMSE) to evaluate the performance of different methods.
The metric MAE is defined as:
\begin{eqnarray}\scriptsize
MAE = \frac{1}{T}\sum_{i,j}|R_{ij}-\hat R_{ij}|
\end{eqnarray}
where $R_{ij}$ is the rating user $i$ gave to item $j$, and $\hat{R}_{ij}$ denotes the rating
user $i$ gave to item $j$ as predicted by a method. Particularly,
$\hat R_{ij}$ can be calculated by $U_{i}V_{j}^{T}$ in our model. Moreover,
$T$ is the number of tested ratings. The metric RMSE is defined as:
\begin{eqnarray}\scriptsize
RMSE = \sqrt{\frac{1}{T}\sum_{i,j}(R_{ij}-\hat R_{ij})^2}
\end{eqnarray}
From the definitions, we can see that a smaller MAE or RMSE means better performance.

\begin{table*}[htbp]\scriptsize
\centering
\caption{\label{comparison}Performance Comparisons on Douban Movie (the baseline of improved performance is PMF)}
\resizebox{0.87\textwidth}{!}{
\begin{tabular}{c|c||c|c|c|c|c|c|c|c|c}
\hline
{Training} & {Metrics} & {UserMean} &{ItemMean} & {PMF} & {SoMF} & {Hete-MF} & {Hete-CF}  & {SimMF-mean} & {SimMF-max} & {SimMF-min}\\
\hline
\hline
\multirow{4}{*}{80\%} & {MAE} & {0.6958} & {0.6476} & {0.6325} &{0.6073}& {0.6221}& {0.6273}   & {0.5974}  & {0.6026} & {\textbf{0.5926}} \\
& {Improve} & {-10.01\%} & {-2.83\%} & {}& {3.99\%} & {1.64\%}& {0.82\%}   & {5.55\%}& {4.73\%} & {6.31\%} \\
\cline{2-11}
& {RMSE} & {0.8846} & {0.8537} & {0.8815} & {0.8283}& {0.8609}& {0.8664}   & {0.7729} & {0.7809} & {\textbf{0.7656}} \\
& {Improve} & {-0.35\%} & {3.15\%} & {} & {6.03\%}& {2.34\%}& {1.71\%}   & {12.32\%} & {11.41\%} & {13.14\%} \\
\hline
\multirow{4}{*}{60\%} & {MAE} & {0.6986} & {0.6557} & {0.6591} & {0.6219}&{0.6490}& {0.6509}   & {0.6060} & {0.6110} & {\textbf{0.6008}} \\
& {Improve} & {-6.00\%} & {0.35\%} & {}& {5.63\%} & {1.53\%}& {1.24\%}   & {8.06\%} & {7.30\%} & {8.85\%} \\
\cline{2-11}
& {RMSE} & {0.8925} & {0.8748} & {0.9281} & {0.8584}& {0.9100}& {0.9118}   & {0.7852} & {0.7927} & {\textbf{0.7772}} \\
& {Improve} & {3.84\%} & {5.75\%} & {}& {7.51\%} & {1.95\%} & {1.76\%}  & {15.40\%} & {14.59\%} & {16.26\%} \\
\hline
\multirow{4}{*}{40\%} & {MAE} & {0.7052} & {0.6733} & {0.7092} & {0.6457}&{0.6933}& {0.7029}   & {0.6186} & {0.6237} & {\textbf{0.6134}} \\
& {Improve} & {0.57\%} & {5.07\%} & {} & {8.96\%}& {2.24\%}& {0.89\%}   & {12.77\%} & {12.06\%} & {13.51\%} \\
\cline{2-11}
& {RMSE} & {0.9085} & {0.9139} & {1.0107} &{0.9034}& {0.9842}& {0.9941}   & {0.8023} & {0.8093} & {\textbf{0.7952}} \\
& {Improve} & {10.11\%} & {9.57\%} & {} & {10.62\%}& {2.62\%}& {1.64\%}   & {20.62\%} & {19.93\%} & {21.32\%} \\
\hline
\multirow{4}{*}{20\%} & {MAE} & {0.7227} & {0.7124} & {0.8367} & {0.6973} & {0.8235}& {0.8302}  & {0.6461} & {0.6509} & {\textbf{0.6417}} \\
& {Improve} & {13.63\%} & {14.85\%} & {} &{16.66\%}& {1.58\%}& {0.78\%}   & {22.78\%} & {22.21\%} & {23.31\%} \\
\cline{2-11}
& {RMSE} & {0.9502} & {1.0006} & {1.2060} & {1.0037}& {1.1838}& {1.1963}  & {0.8388}  & {0.8446} & {\textbf{0.8335}} \\
& {Improve} & {21.21\%} & {17.03\%} & {} &{16.78\%}& {1.84\%}& {0.80\%}  & {30.45\%}  & {29.97\%} & {30.89\%} \\
\cline{1-11}
\hline
\hline
\multicolumn{2}{c||}{Running Time(s)} & {0.5157} & {0.5242} & {1096}& {1385} & {4529} & {7342}   & \multicolumn{3}{|c}{3168} \\
\hline
\end{tabular}}
\end{table*}

\begin{table*}[htbp]\scriptsize
\centering
\caption{\label{comparison}Performance Comparisons on Yelp (the baseline of improved performance is PMF)}
\resizebox{0.87\textwidth}{!}{
\begin{tabular}{c|c||c|c|c|c|c|c|c|c|c}
\hline
{Training} & {Metrics} & {UserMean} &{ItemMean} & {PMF} & {SoMF} & {Hete-MF} &{Hete-CF} & {SimMF-mean} & {SimMF-max} & {SimMF-min}\\
\hline
\hline
\multirow{4}{*}{80\%} & {MAE} & {0.9664} & {0.8952} & {1.2201}& {0.8789} & {0.9307}& {1.2117}   & {0.8292}  & {0.8503} & {\textbf{0.8059}} \\
& {Improve} & {20.79\%} & {26.63\%} & {}&{27.96\%} & {23.72\%}& {0.69\%}    & {32.04\%} & {30.31\%} & {33.95\%} \\
\cline{2-11}
& {RMSE} & {1.3443} & {1.2327} & {1.6479}& {1.1912}& {1.2773} & {1.6249}   & {1.0577}  & {1.0708} & {\textbf{1.0465}} \\
& {Improve} & {18.42\%} & {25.20\%} & {} & {27.71\%}& {22.49\%}& {1.40\%}   & {35.82\%}  & {35.02\%} & {36.49\%} \\
\hline
\multirow{4}{*}{60\%} & {MAE} & {0.9803} & {0.9247} & {1.3835}& {0.9156} & {0.9708}& {1.3510}   & {0.8366} & {0.8615} & {\textbf{0.8109}} \\
& {Improve} & {29.14\%} & {33.16\%} & {}& {33.82\%} & {29.83\%}& {2.35\%}   & {39.53\%} & {37.73\%} & {41.39\%} \\
\cline{2-11}
& {RMSE} & {1.3556} & {1.2893} & {1.8438}&{1.2591} & {1.3352}& {1.7940}   & {1.0684}  & {1.0842} & {\textbf{1.0532}} \\
& {Improve} & {26.48\%} & {30.07\%} & {} & {31.71\%} & {27.58\%}& {2.70\%}  & {42.05\%} & {41.20\%} & {42.88\%} \\
\hline
\multirow{4}{*}{40\%} & {MAE} & {1.0219} & {0.9819} & {1.7081} & {0.9790}& {1.0409}& {1.6360}    & {0.8509} & {0.8810} & {\textbf{0.8186}} \\
& {Improve} & {40.17\%} & {42.52\%} & {}& {42.68\%} & {39.06\%}& {4.22\%}   & {50.18\%} & {48.42\%} & {52.18\%} \\
\cline{2-11}
& {RMSE} & {1.4241} & {1.3873} & {2.2123} & {1.3682}& {1.4343} & {2.1116}  & {1.0863} & {1.1031} & {\textbf{1.0686}} \\
& {Improve} & {35.63\%} & {37.29\%} & {} &{38.15\%} & {35.17\%}& {4.55\%}  & {50.90\%}  & {50.12\%} & {51.70\%} \\
\hline
\multirow{4}{*}{20\%} & {MAE} & {1.1344} & {1.1202} & {2.6935} & {1.1252}& {1.8429}& {2.5782}    & {0.8687} & {0.9047} & {\textbf{0.8290}} \\
& {Improve} & {57.88\%} & {58.41\%} & {} & {58.23\%}& {31.58\%}& {4.28\%}   & {67.75\%}  & {66.41\%} & {69.22\%} \\
\cline{2-11}
& {RMSE} & {1.5958} & {1.5981} & {3.2512} & {1.5907}& {2.3357}& {3.0807}   & {1.1307} & {1.1733} & {\textbf{1.0944}} \\
& {Improve} & {50.92\%} & {50.85\%} & {} & {51.07\%} & {28.16\%}& {5.24\%}  & {65.22\%} & {63.91\%} & {66.34\%} \\
\cline{1-11}
\hline
\hline
\multicolumn{2}{c||}{Running Time(s)} & {0.0646} & {0.0642} & {100}& {137} & {1963}& {2378}   & \multicolumn{3}{|c}{1414} \\
\hline
\end{tabular}}
\end{table*}

\begin{table*}[htbp]\scriptsize
\centering
\caption{\label{comparison}Performance Comparisons on MovieLens (the baseline of improved performance is PMF)}
\resizebox{0.87\textwidth}{!}{
\begin{tabular}{c|c||c|c|c|c|c|c|c|c}
\hline
{Training} & {Metrics} & {UserMean} &{ItemMean} & {PMF} & {Hete-MF} &{Hete-CF} & {SimMF-mean} & {SimMF-max} & {SimMF-min}\\
\hline
\hline
\multirow{4}{*}{80\%} & {MAE} & {0.8439} & {0.7911} & {0.7902} & {0.7659} & {0.8088}  & {0.7491}  & {0.7615} & {\textbf{0.7289}} \\
& {Improve} & {-6.80\%} & {-0.11\%} & {}& {3.08\%} & {-2.35\%}   &{5.20\%} & {3.63\%} & {7.76\%} \\
\cline{2-10}
& {RMSE} & {1.0594} & {0.9961} & {1.0111} & {0.9721}& {1.0366} & {0.9437}  & {0.9559} & {\textbf{0.9215}} \\
& {Improve} & {-4.78\%} & {1.48\%} & {}& {3.86\%} & {-2.52\%}  & {6.67\%} & {5.46\%}  & {8.86\%}  \\
\hline
\multirow{4}{*}{60\%} & {MAE} & {0.8527} & {0.7962} & {0.8252}  & {0.7841}& {0.8644} & {0.7623} & {0.7727} & {\textbf{0.7496}} \\
& {Improve} & {-3.33\%} & {3.51\%} & {} & {4.98\%} & {-4.75\%} & {7.62\%} & {6.36\%} & {9.16\%} \\
\cline{2-10}
& {RMSE} & {1.0753} & {1.0051} & {1.0549} & {0.9971}& {1.1119}   & {0.9592}  & {0.9688} & {\textbf{0.9456}} \\
& {Improve} & {-1.93\%} & {4.72\%} & {} & {5.48\%}& {-5.40\%}  & {9.07\%} & {8.16\%} & {10.36\%}  \\
\hline
\multirow{4}{*}{40\%} & {MAE} & {0.8745} & {0.8062} & {0.8992}  & {0.8283}& {1.0000}  & {0.7790} & {0.7880} & {\textbf{0.7711}} \\
& {Improve} & {2.75\%} & {10.34\%} & {} & {7.88\%} & {-11.21\%} & {13.37\%} & {12.37\%} & {14.25\%} \\
\cline{2-10}
& {RMSE} & {1.1169} & {1.0222} & {1.1526} & {1.0595}& {1.2918}  & {0.9789} & {0.9861} & {\textbf{0.9719}} \\
& {Improve} & {3.10\%} & {11.31\%} & {} & {8.08\%}& {-12.08\%}  &{15.07\%} & {14.45\%}  & {15.68\%} \\
\hline
\multirow{4}{*}{20\%} & {MAE} & {0.9561} & {0.8378} & {1.2942}  & {1.1104}& {1.5824}  & {0.8114} & {0.8154} & {\textbf{0.8139}} \\
& {Improve} & {26.12\%} & {35.27\%} & {} & {14.20\%}& {-22.27\%}  & {37.30\%} & {37.00\%}  & {37.11\%}  \\
\cline{2-10}
& {RMSE} & {1.2724} & {1.0780} & {1.6251} & {1.4280}& {1.9427}   & {1.0156} & {1.0167} & {\textbf{1.0213}} \\
& {Improve} & {21.70\%} & {33.67\%} & {} & {12.13\%}& {-19.54\%}  & {37.51\%} & {37.44\%} & {37.15\%} \\
\cline{1-10}
\hline
\hline
\multicolumn{2}{c||}{Running Time(s)} & {0.0575} & {0.0555} & {80} & {183}& {295}  & \multicolumn{3}{|c}{159} \\
\hline
\end{tabular}}
\end{table*}

\begin{table*}[htbp]\scriptsize
\centering
\caption{\label{tb2}Performance Comparisons on Douban Book (the baseline of improved performances is PMF)}
\resizebox{0.87\textwidth}{!}{
\begin{tabular}{c|c||c|c|c|c|c|c|c|c|c}
\hline
{Training} & {Metrics} & {UserMean} &{ItemMean} & {PMF} & {SoMF} & {Hete-MF} &{Hete-CF} & {SimMF-mean} & {SimMF-max} & {SimMF-min}\\
\hline
\hline
\multirow{4}{*}{80\%} & {MAE} & {0.6204} & {0.6050} & {0.5754} & {0.5748} & {0.5709} & {0.5815}  & {0.5517}  & {0.5540} & {\textbf{0.5495}} \\
& {Improve} & {-7.81\%} & {-5.15\%} & {} & {0.11\%} & {0.79\%} & {-1.06\%}  &{4.11\%} & {3.72\%} & {4.50\%} \\
\cline{2-11}
& {RMSE} & {0.7902} & {0.7588} & {0.7454} & {0.7294} & {0.7309}& {0.7573} & {0.6974}  & {0.7011} & {\textbf{0.6937}} \\
& {Improve} & {-6.01\%} & {-0.79\%} & {} & {2.14\%} & {1.94\%} & {-1.61\%}  & {6.44\%} & {5.94\%}  & {6.93\%}  \\
\hline
\multirow{4}{*}{60\%} & {MAE} & {0.6244} & {0.6090} & {0.6008} & {0.5902} & {0.5822}& {0.6068} & {0.5569} & {0.5594} & {\textbf{0.5543}} \\
& {Improve} & {-3.93\%} & {-1.37\%} & {} & {1.77\%} & {3.10\%} & {-1.00\%} & {7.32\%} & {6.89\%} & {7.74\%} \\
\cline{2-11}
& {RMSE} & {0.7998} & {0.7588} & {0.7827} & {0.7520} &{0.7480}& {0.7953}   & {0.7028}  & {0.7068} & {\textbf{0.6988}} \\
& {Improve} & {-2.19\%} & {3.06\%} & {} & {3.92\%} & {4.43\%}& {-1.61\%}  & {10.21\%} & {9.69\%} & {10.71\%}  \\
\hline
\multirow{4}{*}{40\%} & {MAE} & {0.6325} & {0.6231} & {0.6696} & {0.6141} & {0.6008}& {0.6767}  & {0.5700} & {0.5749} & {\textbf{0.5651}} \\
& {Improve} & {5.55\%} & {6.96\%} & {} & {8.29\%} & {10.28\%} & {-1.05\%} & {14.87\%} & {14.15\%} & {15.62\%} \\
\cline{2-11}
& {RMSE} & {0.8193} & {0.7933} & {0.8885} & {0.7903} & {0.7764}& {0.9027}  & {0.7189} & {0.7277} & {\textbf{0.7102}} \\
& {Improve} & {7.78\%} & {10.72\%} & {} & {11.05\%} & {12.61\%}& {-1.60\%}  &{19.08\%} & {18.10\%}  & {20.06\%} \\
\hline
\multirow{4}{*}{20\%} & {MAE} & {0.6617} & {0.7068} & {0.9873} & {0.6329} & {0.6582}& {1.0695}  & {0.6306} & {0.6439} & {\textbf{0.6174}} \\
& {Improve} & {32.98\%} & {28.41\%} & {} & {35.89\%}& {33.33\%}& {-8.32\%}  & {36.12\%} & {34.79\%}  & {37.47\%}  \\
\cline{2-11}
& {RMSE} & {0.8906} & {1.0033} & {1.3251} & {0.8245} &{0.8660}& {1.4294} & {0.8003} & {0.8260} & {\textbf{0.7746}} \\
& {Improve} & {32.79\%} & {24.29\%} & {} & {37.78\%} & {34.65\%}& {-7.88\%}  & {39.60\%} & {37.67\%} & {41.54\%} \\
\cline{1-11}
\hline
\hline
\multicolumn{2}{c||}{Running Time(s)} & {0.2957} & {0.2797} & {787} & {982} & {1071}& {1147}  & \multicolumn{3}{|c}{1064} \\
\hline
\end{tabular}}
\end{table*}
\subsection{Compared Methods}
In this section, we compare SimMF with six representative methods. There are different variations for SimMF. We use SimMF-U($y$)I($y$)  to represent SimMF with regularization on users and items, where $y\in\{a,i\}$ and it represents the average or individual based regularization. Similarly, SimMF-U($y$) (SimMF-I($y$)) means SimMF with regularization only on users (items). There are six baseline methods, including four types. There are two basic methods (i.e., UserMean and ItemMean), a collaborative filtering with low-rank matrix factorization (i.e., PMF), a social recommendation method (i.e., SoMF) and two HIN based methods (i.e., Hete-MF and Hete-CF). These baselines are summarized as follows.

{
\textbullet\ UserMean. This method uses the mean value of every user to predict the missing values.

\textbullet\ ItemMean. This method utilizes the mean value of every item to predict the missing values.

\textbullet\ PMF. This method is a typical matrix factorization method proposed by Salakhutdinov and Minh \cite{SM95}. And in fact it is equivalent to basic low-rank matrix factorization in Section 4.1.

\textbullet\ SoMF. This is the matrix factorization based recommendation method with social average-based regularization proposed by Ma et al. \cite{MZLLK11}.

\textbullet\ Hete-MF. This is the matrix factorization based recommendation framework combining user ratings and various entity similarity matrices proposed by Yu et al. \cite{YRGSH13}.

\textbullet\ Hete-CF. Luo et al. proposed the social collaborative filtering algorithm using heterogeneous relations \cite{LPW14}.
} \\

We employ HeteSim \cite{SKHYW14} to evaluate the similarity of objects. For the Douban Movie dataset, we use 7 meaningful meta paths for user whose length is smaller than 4 (i.e., UU, UGU, ULU, UMU, UMDMU, UMTMU, UMAMU) and 5 meaningful meta paths for movie whose length is smaller than 3 (i.e., MTM, MDM, MAM, MUM, MUUM). For the Yelp dataset, we use 4 meta paths for user (i.e., UU, UBU, UBCBU, UBLBU) and 4 meta paths for business (i.e., BUB, BCB, BLB, BUUB). Similarly, we utilize 5 meta paths for user (i.e., UGU, UAU, UOU, UMU, UMTMU) and 2 meta paths for movie (i.e., MTM, MUM) for the MovieLens dataset. And for the Douban Book dataset, we utilize 7 meta paths for user (i.e., UU, UGU, ULU, UBU, UBABU, UBPBU, UBYBU) and 5 meta paths for book (i.e., BAB, BPB, BYB, BUB, BUUB). These similarity data are fairly used for Hete-CF and SimMF. Hete-MF uses similarity data of users, since the model only considers the similarity relationships between items.

\begin{figure*}[htbp]\scriptsize
\subfigure[Douban Movie, MAE]{
\begin{minipage}[t]{0.3\linewidth}
\centering
\includegraphics[width=4.4cm]{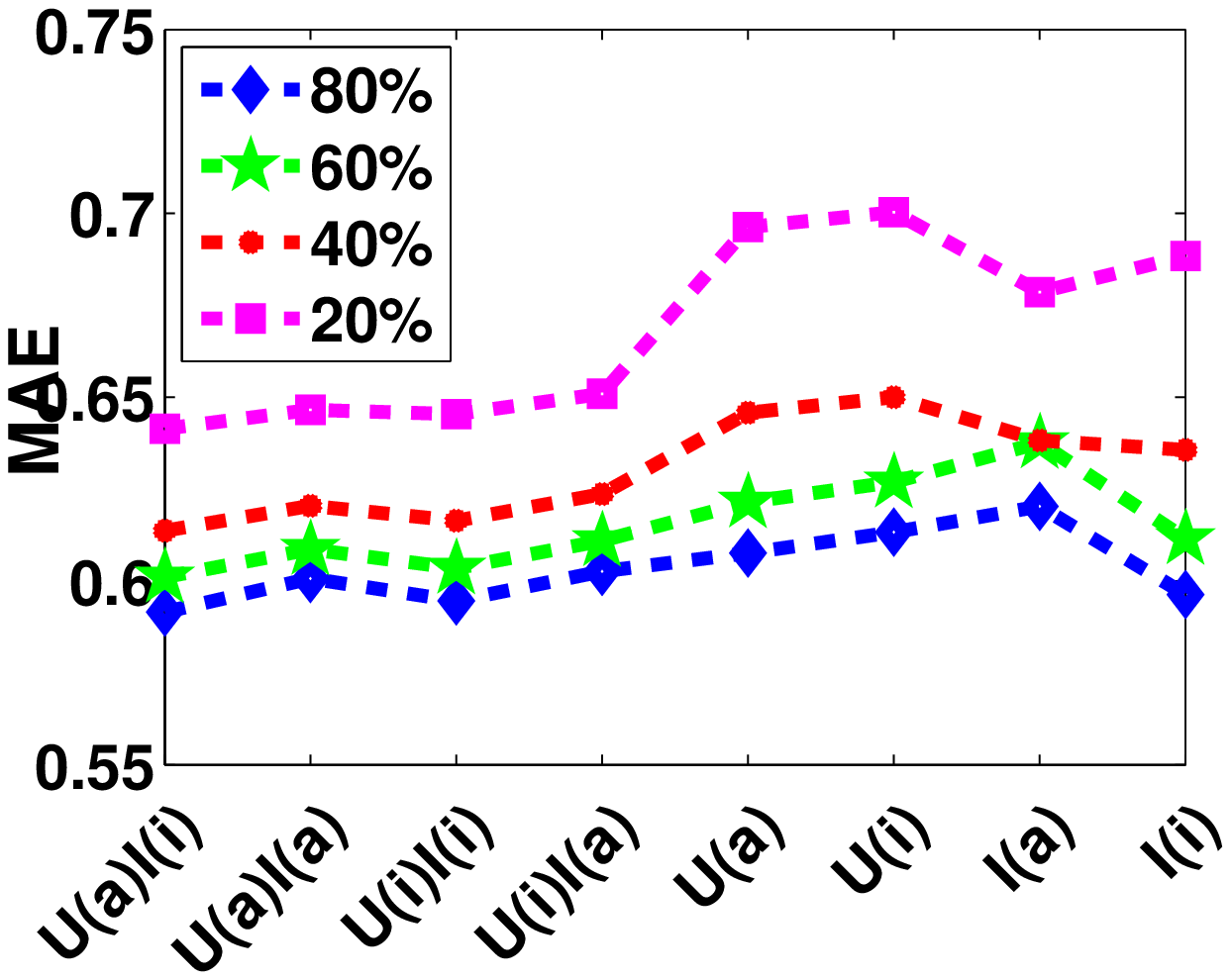}
\end{minipage}%
}
\hspace{-50pt}
\subfigure[Douban Movie, RMSE]{
\begin{minipage}[t]{0.3\linewidth}
\centering
\includegraphics[width=4.4cm]{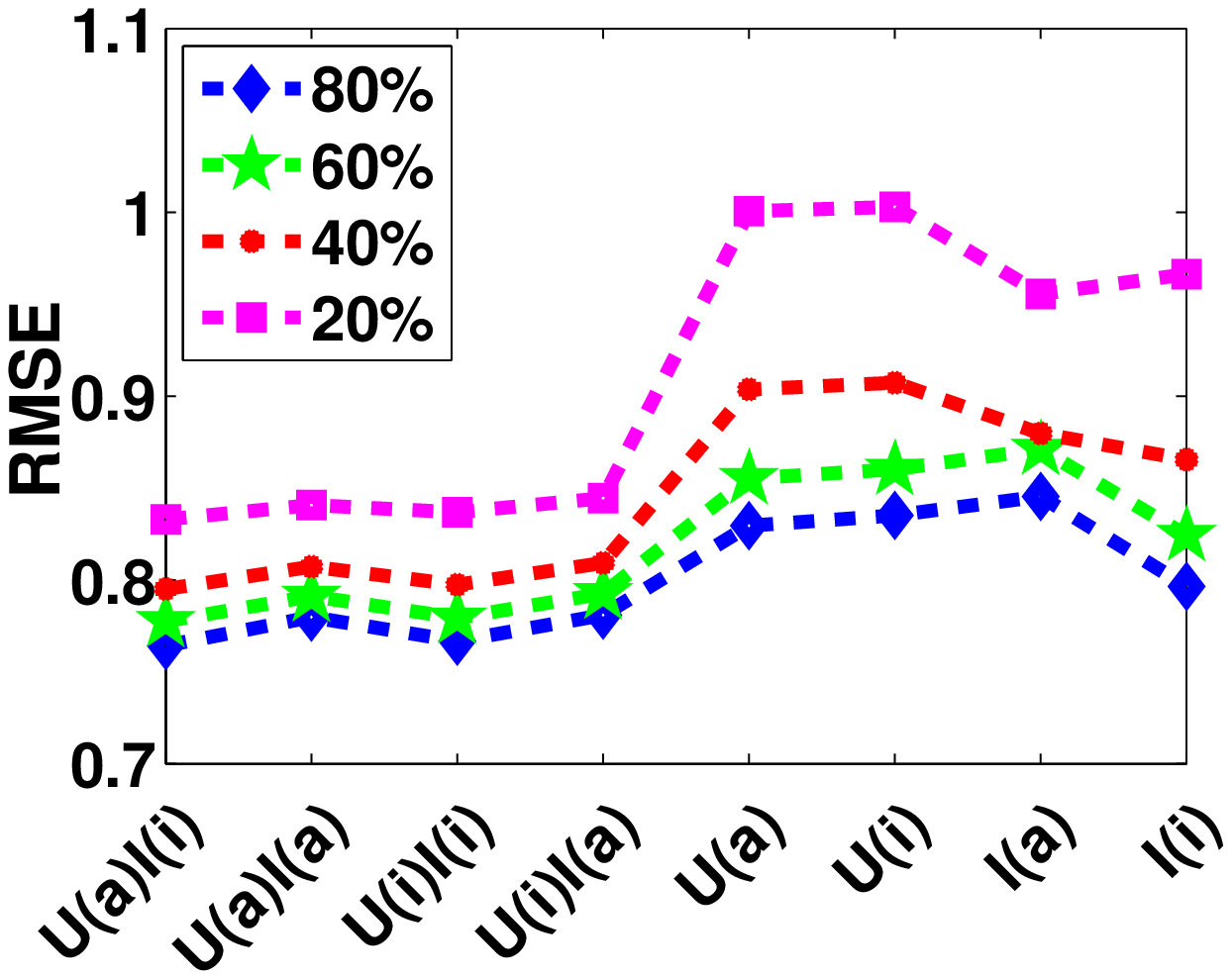}
\end{minipage}%
}
\hspace{-50pt}
\subfigure[Yelp, MAE]{
\begin{minipage}[t]{0.3\linewidth}
\centering
\includegraphics[width=4.4cm]{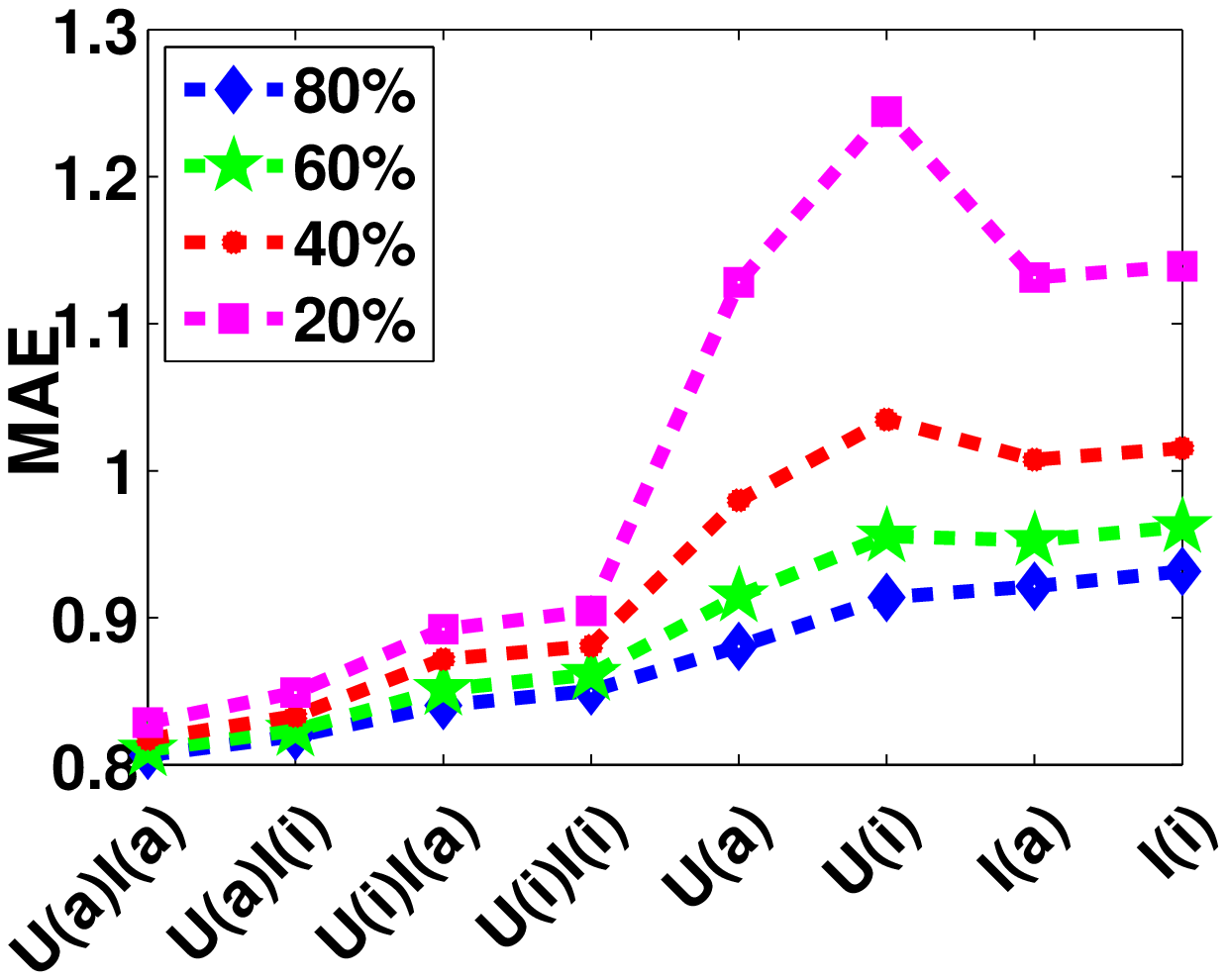}
\end{minipage}%
}
\hspace{-50pt}
\subfigure[Yelp, RMSE]{
\begin{minipage}[t]{0.3\linewidth}
\centering
\includegraphics[width=4.4cm]{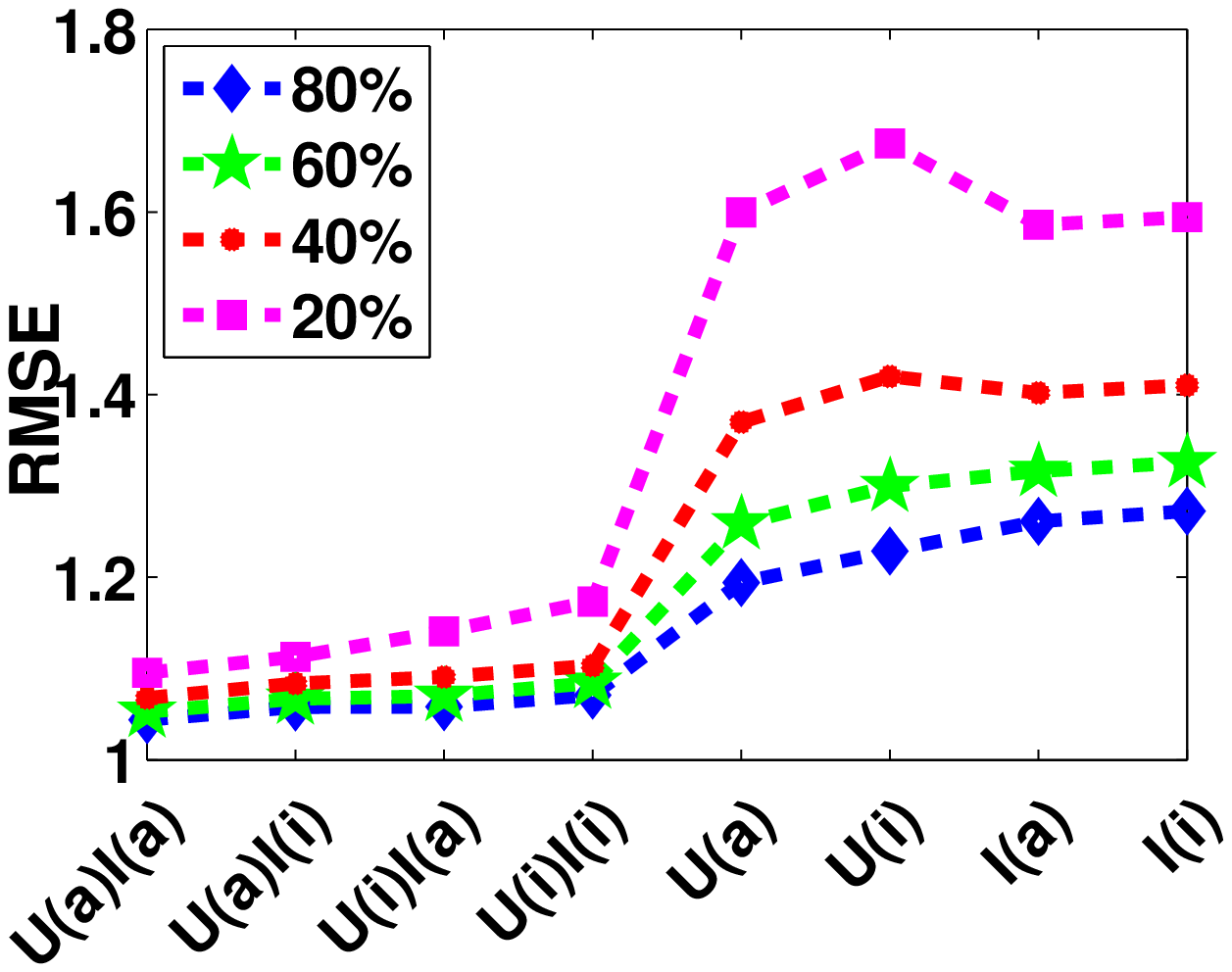}
\end{minipage}%
}
\caption{performance of SimMF with different regularizations on Douban Movie and Yelp datasets.}
\end{figure*}
\subsection{Effectiveness Experiments}
This section will validate the effectiveness of SimMF through comparing its different variations to baselines. Here we run four versions of SimMF-U($y$)I($y$) ($y\in\{a,i\}$), and record the worst (denoted as SimMF-max in Tables 2-5), the best (denoted as SimMF-min) and average (denoted as SimMF-mean) performance of these four versions. The $\alpha$ and $\beta$ are set to 100 and 10 respectively for Douban Movie dataset, as suggested in the following parameter experiment. For other datasets, $\alpha$ and $\beta$ are set to the optimal values according to related parameter experiments. For all the experiments in this paper, the values of $\lambda_{1}$ and $\lambda_{2}$ are set to a trivial value 0.001 and the length of latent feature vectors $U_{i}$ and $V_{j}$ are set to 10. The parameters of other methods are set to the optimal values obtained in parameter experiments.

For these datasets, we use different ratios (80\%, 60\%, 40\%, 20\%) of data as training set. For example, the training data 80\% means that we select 80\% of the ratings from user-item rating matrix as the training data to predict the remaining 20\% of ratings. The random selection was carried out 10 times independently in all the experiments. We report average results on Douban Movie, Yelp, MovieLens and Douban Book datasets in Tables 2, 3, 4, 5 respectively and record the improvement ratio of all methods compared to the PMF. In addition, we also report the average running time of these methods with the 80\% training ratio in the last line of above tables. For those HIN based methods (i.e., Hete-CF, Hete-MF, and SimMF), we only report the running time of the model learning process, ignoring the running time of similarity computation. Note that, we report the mean running time for SimMF, since the four versions of SimMF have the similar computational complexity.

The results are shown in Tables 2-5. Note that SoMF is absent in Table 4 because there is no social relation in MovieLens dataset. From the experimental comparisons, we can observe the following phenomena.
\begin{itemize}
\item The SimMF always outperforms the baselines in most conditions, even for the worst performance of SimMF (i.e., SimMF-max). It validates that more attribute information from users and items exploited in SimMF is really helpful to improve the recommendation performance. In addition, the model integrating more information usually has better performances. That is the reason why other matrix factorization models integrating heterogeneous information usually have better performance than the basic matrix factorization model PMF.
\item Although Hete-MF and Hete-CF also utilize the attribute information from users and items, they have worse performance than SimMF, which implies the proposed SimMF has better mechanism to integrate heterogeneous information. We know that Hete-MF only integrates attribute information of items, while the same parameter for similarity regularization terms of users and items may cause the bad performance of Hete-CF.
\item When considering different training data ratios, we can find that the superiority of SimMF is more significant for less training data. It indicates that SimMF can effectively alleviate data sparsity problem. We think the reason lies in that, through exploiting different meta paths, we can make full use of rich attribute information of users and items to reflect the similarity of users and items from different aspects. The integration of similarities can comprehensively reveal the similarity of users and items, which compensates for shortage of training data.
\end{itemize}

In addition, we conduct the $t$-test experiments with 95 percent confidence and record the $p$-values of all other models' MAE and RMSE compared to those of PMF, then report them in Tables \ref{ttest1} to \ref{ttest4}. Through checking these tables, the $p$-values of SimMF-max, SimMF-mean, SimMF-min on MAE/RMSE are all less than 5\%. Thus, we can conclude that the $t$-test with 95 percent confidence shows the improvement on MAE and RMSE is statistically stable and non-contingent.

Comparing results of PMF on these four datasets, we can find the performances of PMF are greatly affected by the
density of rating matrix. For Douban Movie (see Table 2) and Douban Book (see Table 5) datasets, PMF performs reasonably, while its performance degrades greatly on Yelp dataset (see Table 3) because of the very sparse rating data on Yelp dataset. When comparing results of SoMF to PMF, it marginally improves the performance on Douban Movie dataset because of the sparse social relations on Douban Movie, while it obviously improves the performance on Yelp dataset due to the sparse social relations on Yelp. So we can conclude that the recommendation performance of SoMF is largely affected by the density of social relations. However, no matter how dense or sparse rating and social relations, SimMF can always achieves the best performance through making full use of the rich attribute information.

Observing the running time of different methods in the last row of Table 2-5, we can find that the running time becomes longer as the models become more complex. That is, HIN based methods (i.e., Hete-MF, Hete-CF, and SimMF) have longer running time than the other methods, since they have more parameters to be learned. However, SimMF is still faster than the other two HIN based methods because SimMF does not need to learn the weights of meta paths.

\begin{table*}[htbp]\scriptsize
\centering
\caption{\label{ttest1}Test of statistical significance on Douban Movie(compared to PMF)}
\resizebox{0.9\textwidth}{!}{
\begin{tabular}{c|c||c|c|c|c|c|c|c|c}
\hline
{Training} & {$p$ Value on} & {UserMean} &{ItemMean} & {SoMF} & {Hete-MF} & {Hete-CF} & {SimMF-mean} & {SimMF-max} & {SimMF-min}\\
\hline
\hline
\multirow{2}{*}{80\%} & {mae} & {1.0352e-06} & {3.0554e-04}&{4.0615e-05}&	{1.2893e-03}	& {1.5761e-02} & {1.1789e-07}	&{5.2338e-05}&{1.6648e-05}\\
\cline{2-10}
& {rmse} & {2.9613e-01}&	{6.5016e-04}&	{5.1011e-05}&	{1.3376e-03}&	{3.1635e-03}	&{2.7626e-07}&{4.0459e-06}	&{1.4730e-06}\\
\hline
\multirow{2}{*}{60\%} & {mae} &{2.7000e-05}	&{1.3604e-01}&{1.0961e-04}&{5.2177e-03}	&{9.5193e-03}	&{2.1732e-07}		&{5.5110e-05}	&{5.7332e-06}\\
\cline{2-10}
& {rmse} & {4.0788e-05}	&{1.3819e-05}&{4.7455e-06}&{9.6322e-04}	&{7.4915e-04}		&{1.0723e-06}	&{3.3469e-07}	&{2.9857e-07}\\
\hline
\multirow{2}{*}{40\%} & {mae} & {2.8287e-02}	&{7.1967e-06}&{6.8801e-06}	&{2.4970e-04}&{8.1609e-03}		&{3.8237e-08}&{3.1140e-07}	&{1.9763e-07}\\
\cline{2-10}
& {rmse} & {6.0979e-07}&	{1.2793e-06}&	{3.1282e-06}&	{2.1936e-04}&	{1.3401e-03} 	&{2.7470e-06}&{6.8430e-08}&	{5.2208e-08}\\
\hline
\multirow{2}{*}{20\%} & {mae} & {1.7764e-05}	&{1.8808e-06}&{1.4798e-06}&{1.0958e-02}	&{1.5353e-01}		&{9.7992e-04}	&{4.6958e-07}	&{3.4183e-07}\\
\cline{2-10}
& {rmse} & {2.1289e-07}&	{5.1176e-07}&{5.4382e-07}&	{3.6990e-03}&	{8.0264e-02}	&{2.7220e-05}	&{7.9077e-08}	&{5.5366e-08}\\
\hline
\end{tabular}}
\end{table*}

\begin{table*}[htbp]\scriptsize
\centering
\caption{\label{ttest2}Test of statistical significance on Yelp (compared to PMF)}
\resizebox{0.9\textwidth}{!}{
\begin{tabular}{c|c||c|c|c|c|c|c|c|c}
\hline
{Training} & {$p$ Value on} & {UserMean} &{ItemMean} & {SoMF}&{Hete-MF}& {Hete-CF}   &{SimMF-mean} & {SimMF-max} & {SimMF-min}\\
\hline
\hline
\multirow{2}{*}{80\%} & {mae} & {1.6087e-08}&	{5.9816e-09}&	{3.0729e-08}&	{2.2094e-08}&	{2.6553e-02}	&{1.7839e-08}&{3.5643e-09}	&{2.2647e-09}\\
\cline{2-10}
& {rmse} & {4.9999e-08}	&{1.6652e-08}&{5.6985e-08}	& {4.0833e-08}	&{3.3981e-03}	&{3.0073e-08}&{5.8683e-09}	&{4.2852e-09}\\
\hline
\multirow{2}{*}{60\%} & {mae} & {1.9104e-05}	&{1.1679e-05}&{1.0755e-05}	&{1.7764e-05}&{6.1223e-02}		&{6.3155e-06}	&{7.1557e-06}	&{4.9394e-06}\\
\cline{2-10}
& {rmse} & {9.5105e-06}	&{5.7887e-06}&{4.6854e-06}	&{8.1686e-06}&{4.6397e-02}			&{1.7077e-06}&{1.7969e-06}	&{1.4553e-06}\\
\hline
\multirow{2}{*}{40\%} & {mae} & {2.5120e-06}& {1.9612e-06}&{1.9719e-06}	&{4.2340e-06}&{1.6990e-02}		&{9.9923e-07}	&{1.1531e-06}	&{8.6473e-07}\\
\cline{2-10}
& {rmse} & {3.0181e-06}	&{2.1545e-06}	&{1.0180e-02}	&{4.7765e-06}&{2.0284e-06}		&{6.5800e-07}&{6.6187e-07}	&{5.8444e-07}\\
\hline
\multirow{2}{*}{20\%} & {mae} & {2.9167e-07}	&{2.8103e-07}&{2.8737e-07}	&{5.0630e-06}&{1.6209e-02}		&{1.6629e-07}	&{1.6671e-07}	&{1.4125e-07}\\
\cline{2-10}
& {rmse} & {8.8131e-06}	&{8.8832e-06}	&{8.7839e-06}&{9.6587e-05}&{1.0616e-01}			&{3.2814e-06}&{3.5585e-06}	&{3.0668e-06}\\
\hline
\end{tabular}}
\end{table*}

\begin{table*}[htbp]\scriptsize
\centering
\caption{\label{ttest3}Test of statistical significance on MovieLens (compared to PMF)}
\resizebox{0.9\textwidth}{!}{
\begin{tabular}{c|c||c|c|c|c|c|c|c}
\hline
{Training} & {$p$ Value on} & {UserMean} &{ItemMean} &{Hete-MF}& {Hete-CF} & {SimMF-mean} &{SimMF-max} & {SimMF-min}\\
\hline
\hline
\multirow{2}{*}{80\%} & {mae} &{9.4604e-06}	&{4.0891e-01}&{2.6496e-05}	&{3.8850e-05}	&{1.4000e-03}	&{1.6748e-06}	&{2.5420e-05} \\
\cline{2-9}
& {rmse} & {1.0173e-02}	&{4.4568e-04}	&{1.1103e-04}&{2.0969e-04}		&{4.4006e-06}&{7.2472e-07}	&{1.5401e-05} \\
\hline
\multirow{2}{*}{60\%} & {mae} & {3.2153e-05}	&{1.0786e-05}	&{1.4067e-05}&{4.1591e-05}		&{1.3101e-06}&{9.7263e-08}	&{8.6432e-07} \\
\cline{2-9}
& {rmse} & {2.0902e-01}	&{8.5753e-06}	&{2.8230e-05}&{9.3957e-05}		&{6.0737e-05}&{3.4647e-07}	&{2.0336e-06}\\
\hline
\multirow{2}{*}{40\%} & {mae} & {1.1104e-03}	&{2.9612e-06}	&{5.0748e-06}&{8.1715e-06}		&{2.7688e-05}&{1.1272e-06}	&{1.0471e-06}\\
\cline{2-9}
& {rmse} & {6.2063e-04}	&{2.2539e-06}	&{4.4253e-06}&{4.1675e-06}	&{5.2865e-06}	&{8.5660e-07}	&{4.1445e-07}\\
\hline
\multirow{2}{*}{20\%} & {mae} & {3.7329e-06}	&{1.0108e-06}	&{1.5296e-04}&{4.2595e-05}		&{8.0936e-07}&{8.3693e-07}	&{8.3246e-07}\\
\cline{2-9}
& {rmse} & {5.2889e-06}	&{6.8341e-07}	&{2.3089e-04}&{3.2462e-05}		&{5.1846e-07}&{4.6419e-07}	&{4.4643e-07}\\
\hline
\end{tabular}}
\end{table*}

\begin{table*}[htbp]\scriptsize
\centering
\caption{\label{ttest4}Test of statistical significance on Douban Book (compared to PMF)}
\resizebox{0.9\textwidth}{!}{
\begin{tabular}{c|c||c|c|c|c|c|c|c|c}
\hline
{Training} & {$p$ Value on} & {UserMean} &{ItemMean} & {SoMF} & {Hete-MF}& {Hete-CF} & {SimMF-mean} & {SimMF-max} & {SimMF-min}\\
\hline
\hline
\multirow{2}{*}{80\%} & {mae} & {1.0531e-06} & {3.2689e-06} & {4.7127e-01} & {6.4948e-03}	& {5.3746e-03} & {6.0729e-06} & {5.9814e-06} & {4.0055e-06}\\
\cline{2-10}
& {rmse} & {3.1196e-06}&	{1.4986e-03}&	{3.0243e-05}&	{1.5466e-04}&	{2.9626e-04} &{4.8773e-07}&{7.3016e-07}	&{3.8479e-07}\\
\hline
\multirow{2}{*}{60\%} & {mae} &{4.8313e-05}	&{1.1407e-03}&{3.5440e-04}&{6.1376e-05}	&{4.7155e-03}	&{1.3429e-06} &{1.6777e-06}	&{1.0946e-06}\\
\cline{2-10}
& {rmse} & {5.1700e-04}	&{5.3944e-05}&{1.2894e-05}&{2.3976e-05}	&{1.1205e-03}		&{3.1942e-07}	&{3.8045e-07}	&{2.7193e-07}\\
\hline
\multirow{2}{*}{40\%} & {mae} & {5.3851e-07}	&{4.5840e-07}&{1.0570e-06}	&{2.0915e-09}&{2.0629e-03}&{4.2202e-10}		&{5.7842e-10}&{3.5349e-10}\\
\cline{2-10}
& {rmse} & {4.7058e-07}&	{9.3940e-07}&	{9.9867e-07}&	{1.5817e-08}&	{2.2916e-04} 	&{1.7519e-09}&{2.2368e-09}&	{1.4566e-09}\\
\hline
\multirow{2}{*}{20\%} & {mae} & {7.5473e-08}	&{1.4155e-07}&{5.4512e-08}&{7.5258e-08}	&{4.139e-05}		&{5.1352e-08}	&{6.0830e-08}	&{4.3955e-08}\\
\cline{2-10}
& {rmse} & {5.8404e-08}&	{2.2978e-07}&{2.9941e-08}&	{5.8076e-08}&	{2.7220e-05}	&{2.5504e-08}	&{3.3022e-08}	&{2.0274e-08}\\
\hline
\end{tabular}}
\end{table*}

\begin{figure*}[htbp]\scriptsize
\subfigure[Paths on users, MAE]{
\begin{minipage}[t]{0.3\linewidth}
\centering
\includegraphics[width=4.2cm]{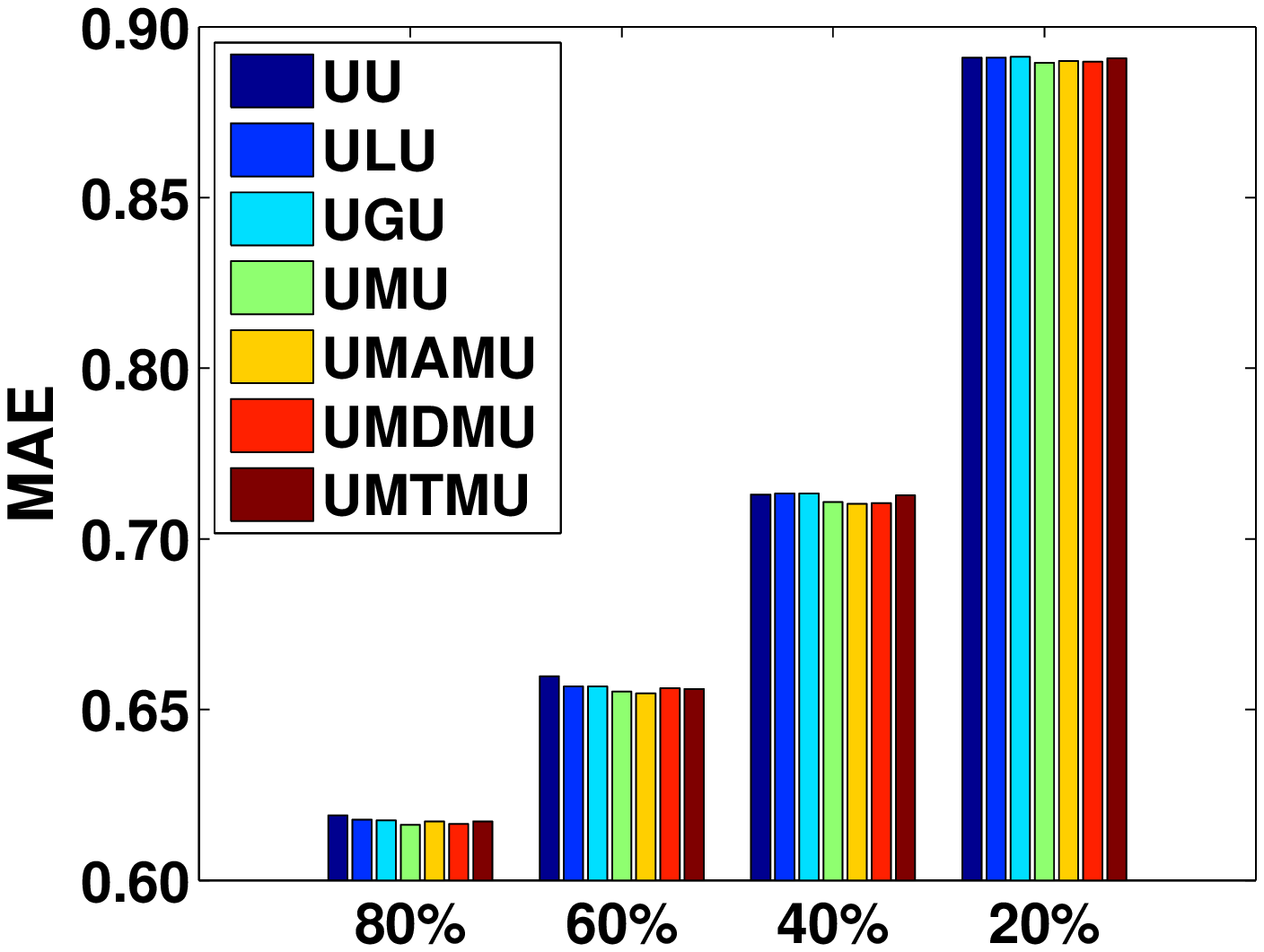}
\end{minipage}%
}
\hspace{-45pt}
\subfigure[Paths on users,  RMSE]{
\begin{minipage}[t]{0.3\linewidth}
\centering
\includegraphics[width=4.2cm]{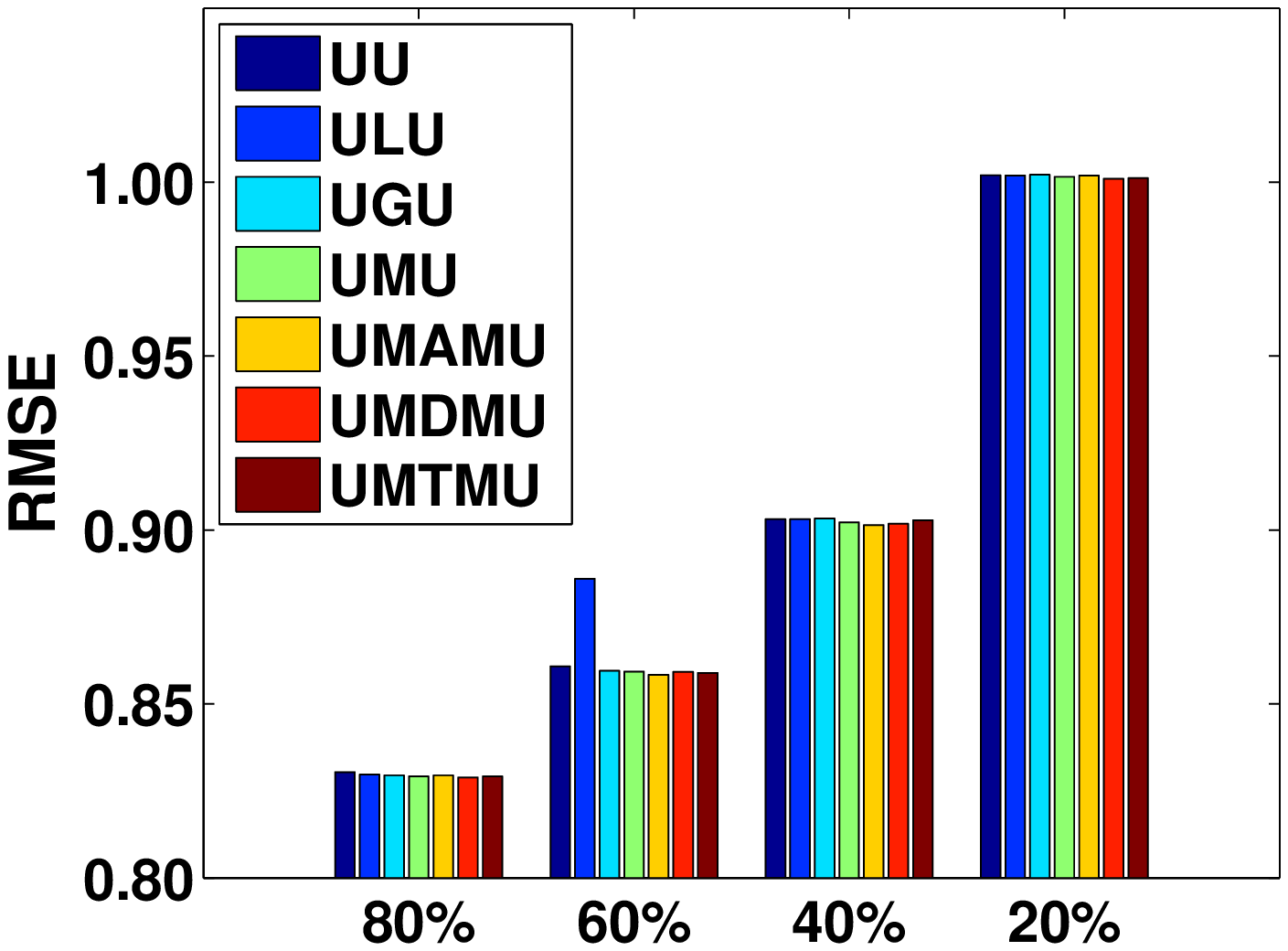}
\end{minipage}%
}
\hspace{-45pt}
\subfigure[Paths on movies, MAE]{
\begin{minipage}[t]{0.3\linewidth}
\centering
\includegraphics[width=4.2cm]{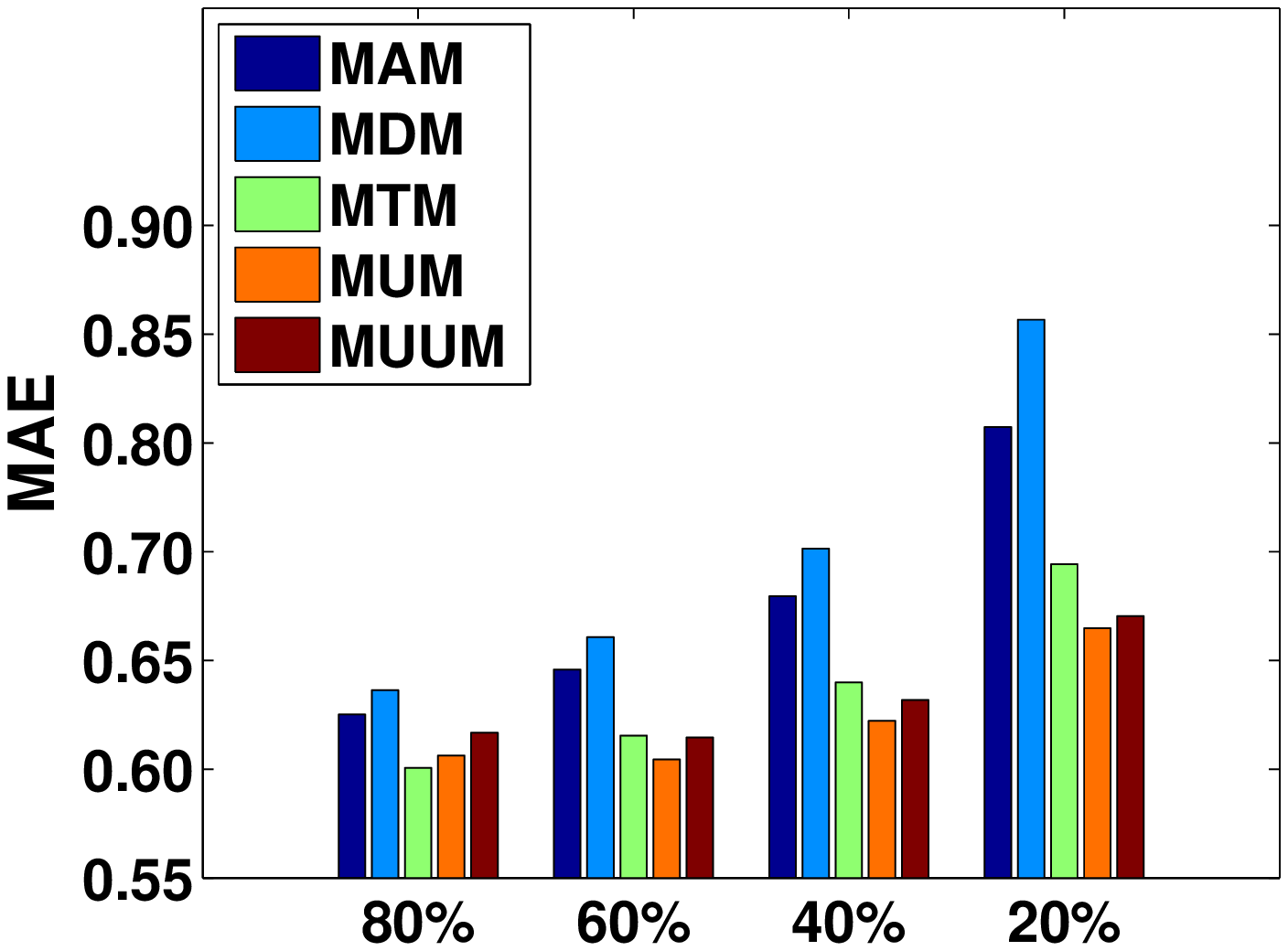}
\end{minipage}%
}
\hspace{-45pt}
\subfigure[Paths on movies, RMSE]{
\begin{minipage}[t]{0.3\linewidth}
\centering
\includegraphics[width=4.2cm]{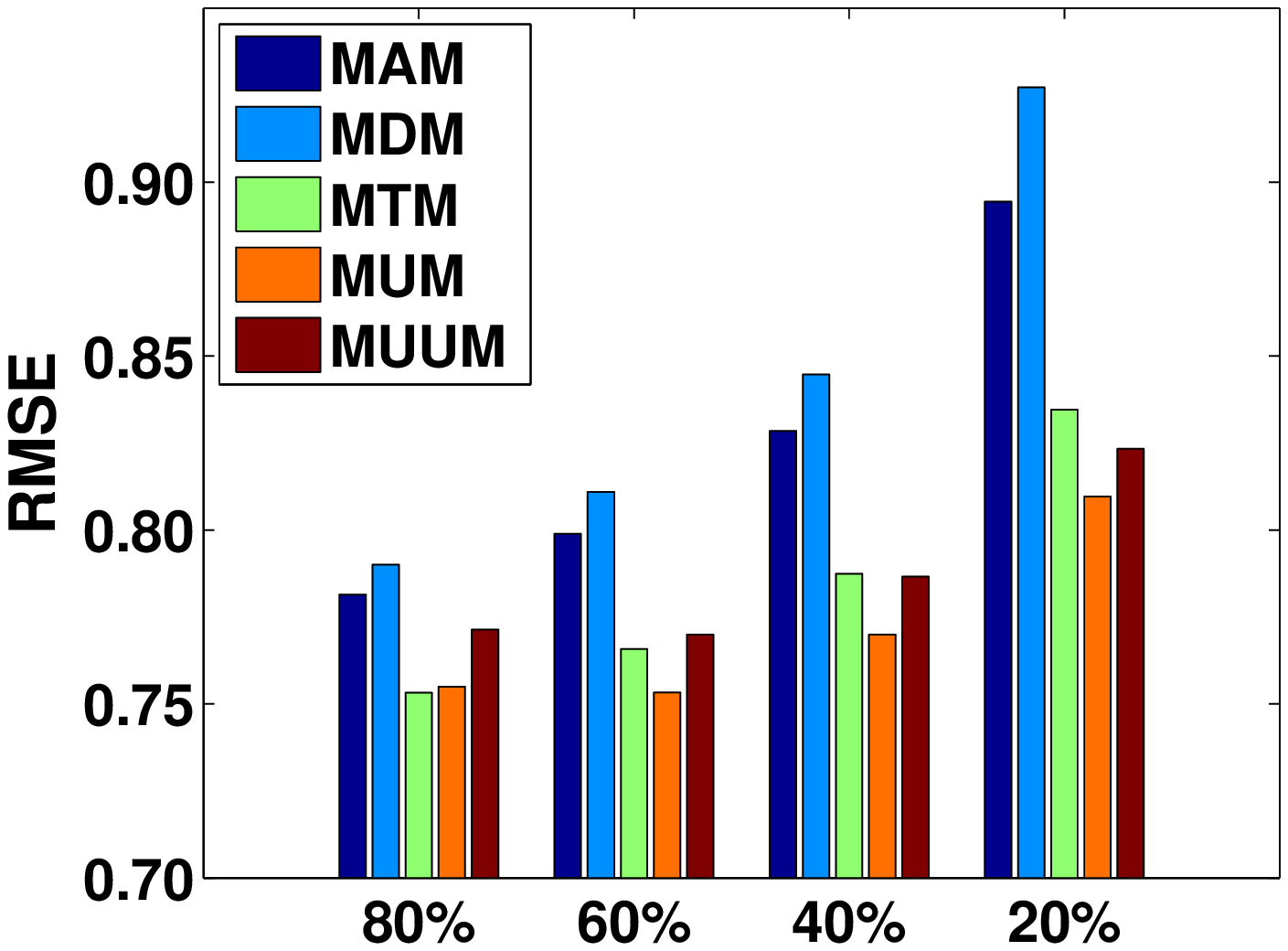}
\end{minipage}%
}
\caption{performance of SimMF with different meta paths on Douban Movie dataset.}
\end{figure*}

\subsection{Impact of Different Regularizations}
Experiments in this section will validate the effect of different regularization models on users and items. Ma et al. \cite{MZLLK11} have explored the effect of average and individual based regularization on social relations of users. However, in this paper, we not only explore the effect on more complex relations, but also consider the effect on both users and items.

We employ four variations of SimMF with average and individual based regularization on users and items (i.e., SimMF with U(a)I(i), U(a)I(a), U(i)I(i), and U(i)I(a)) and four variations of SimMF with average or individual based regularization on users or items (i.e., SimMF with U(a), U(i), I(a), and I(i)). The same parameters are set with above experiments, and the average results are shown in Figure 2. We can find that SimMF integrating similarity information on both users and items always has better performance than it only integrating similarity information on users or items. Again we can observe the difference is far more pronounced when the fraction of training set is low, e.g. at 20\%  SimMF-U(i) and SimMF-U(a) perform very bad.  Moreover, we can also observe an interesting phenomena: regularization models have different effects on users and items.  SimMF-U(a) has better performance than SimMF-U(i) on both datasets, which indicates average-based regularization may be more suitable for users. However, it is not the case for items. SimMF-I(i) performs better than SimMF-I(a) on Douban Movie, while SimMF-I(a) outperforms SimMF-I(i) on Yelp. As a result, SimMF-U(a)I(i) has the best performance on Douban Movie, while SimMF-U(a)I(a) is the best one on Yelp. Although it is hard to draw general conclusions, the above study indicates that different regularization model may significantly affect performance of matrix factorization methods. In summary, we need to find the optimal regularization model according to data properties in real applications.

\subsection{Impact of Different Meta Paths}
In this section, we study the impact of different meta paths. Due to similar analysis, we only show results on Douban Movie dataset. As illustrated above, we employ 7 meta paths on user and 5 meta paths on movie. We will observe performance of SimMF with similarity matrix generated by one single meta path. Under same parameters with above experiments, we run SimMF-U(a) with similarity matrix generated by each meta path on users. Similarly, we also run SimMF-I(i) with similarity matrix generated by each meta path on movies.

The experiment results on Douban Movie dataset are shown in Figure 3. We can observe different impacts of meta paths on users and movies. The SimMF-U(a) with different meta paths (see Figure 3(a,b)) on users all have close performance. Moreover, SimMF-U(a) with MUM has slightly better performance and SimMF-U(a) with UU has worse performance. However, it is not the case for meta paths on items. The SimMF-I(i) with different meta paths on items (see Figure 3(c,d)) have totally different performance. We can find that SimMF-I(i) with MDM has the worst performance, even worse than PMF in some conditions, while SimMF-I(i) with MTM and MUM achieve much better performance on both criteria. We think there are two reasons. (1) Observing Table 1, we can find that the performance of SimMF are much affected by the density of relations. The density of relations on MT and MU are much higher than that on MD and MA. The dense relations are helpful to generate good similarity of items. The similar phenomena have been widely observed in social recommendation \cite{MYLK08,MZLLK11}. (2) The meaningful meta paths are helpful to reveal the similarity of objects. MTM means movies with same type and MUM means movies seen by same users. These two paths are highly correlated as both reveal properties of the movies. These two reasons can also explain the slightly worse performance of the meaningful but sparse UU meta path as compared to other meta paths of users. The experiments imply that we only need to use one single dense and meaningful meta path to generate similarity information, which also can obtain good enough performance.

We further design an experiment to illustrate different importance of meta paths. Concretely, we observe the performance of above SimMF-I(i) with different weight combination methods on 5 meta paths. Except mean weight and random weight on 5 paths, we design a heuristic weight method, i.e., setting the weights according to performance of these paths. That is, paths with good performance have higher weights. Assume MAE performance value of a path ($\mathcal{P}_l$) is $P_l$, and the max MAE value is $P_{max}$. Then the difference is $d_l=e^{P_{max}-P_l}$. And thus the weight of the path is $w^{\mathcal{I}}_l=\frac{d_l}{\sum_l{d_l}}$. The experiment also includes PMF as the baseline. The results are shown in Figure 4. It is obvious that SimMF-I(i) with the heuristic weight method has the best performance, which further validates the meaningful and dense meta paths are more important.

\begin{figure}[htbp]
\centering
\small
\subfigure[MAE]{\label{fig:fft:a}
\begin{minipage}[t]{0.20\textwidth}
  \includegraphics[width=4cm]{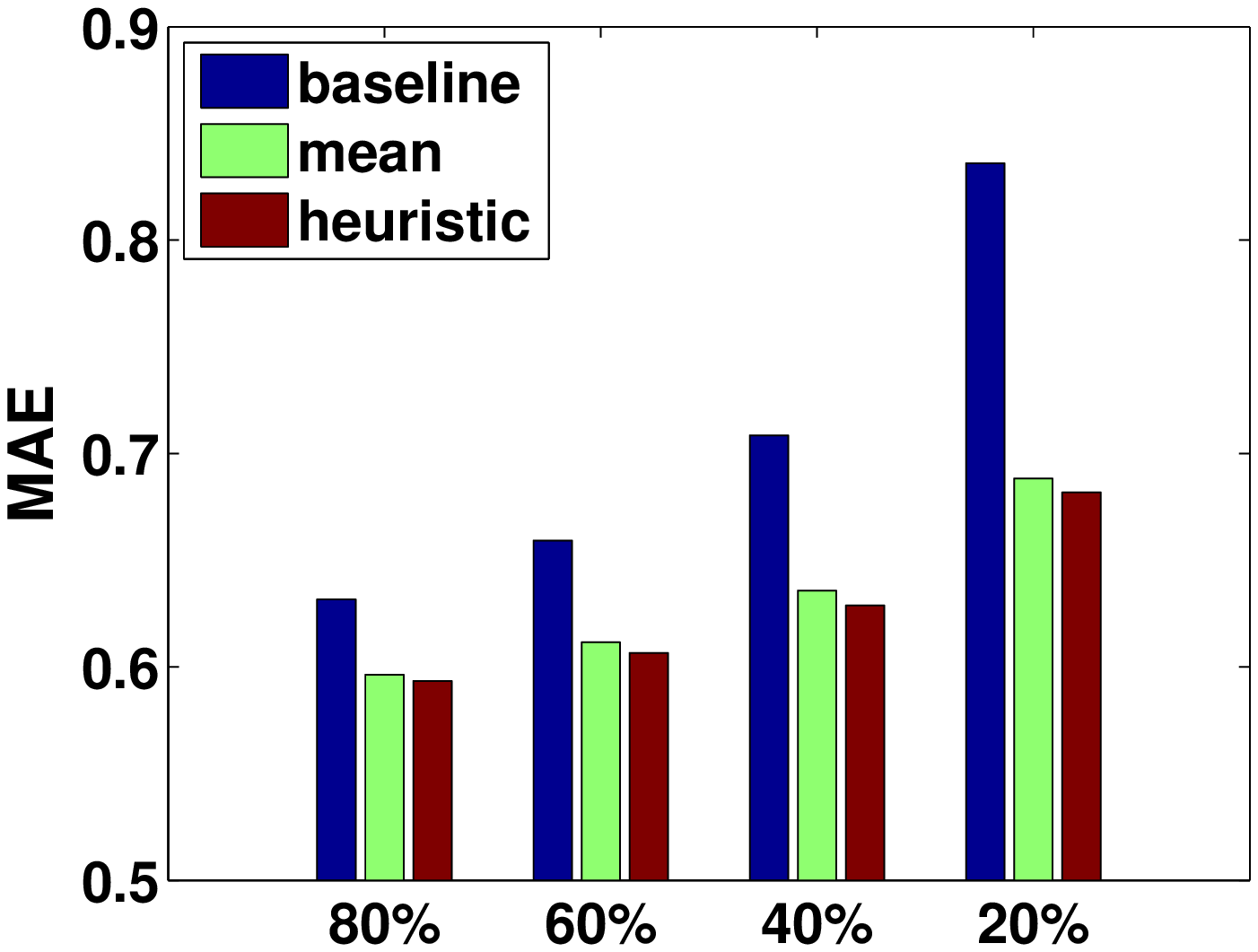}
\end{minipage}%
}
\hspace{0pt}
\subfigure[RMSE]{
\begin{minipage}[t]{0.20\textwidth}
  \includegraphics[width=4cm]{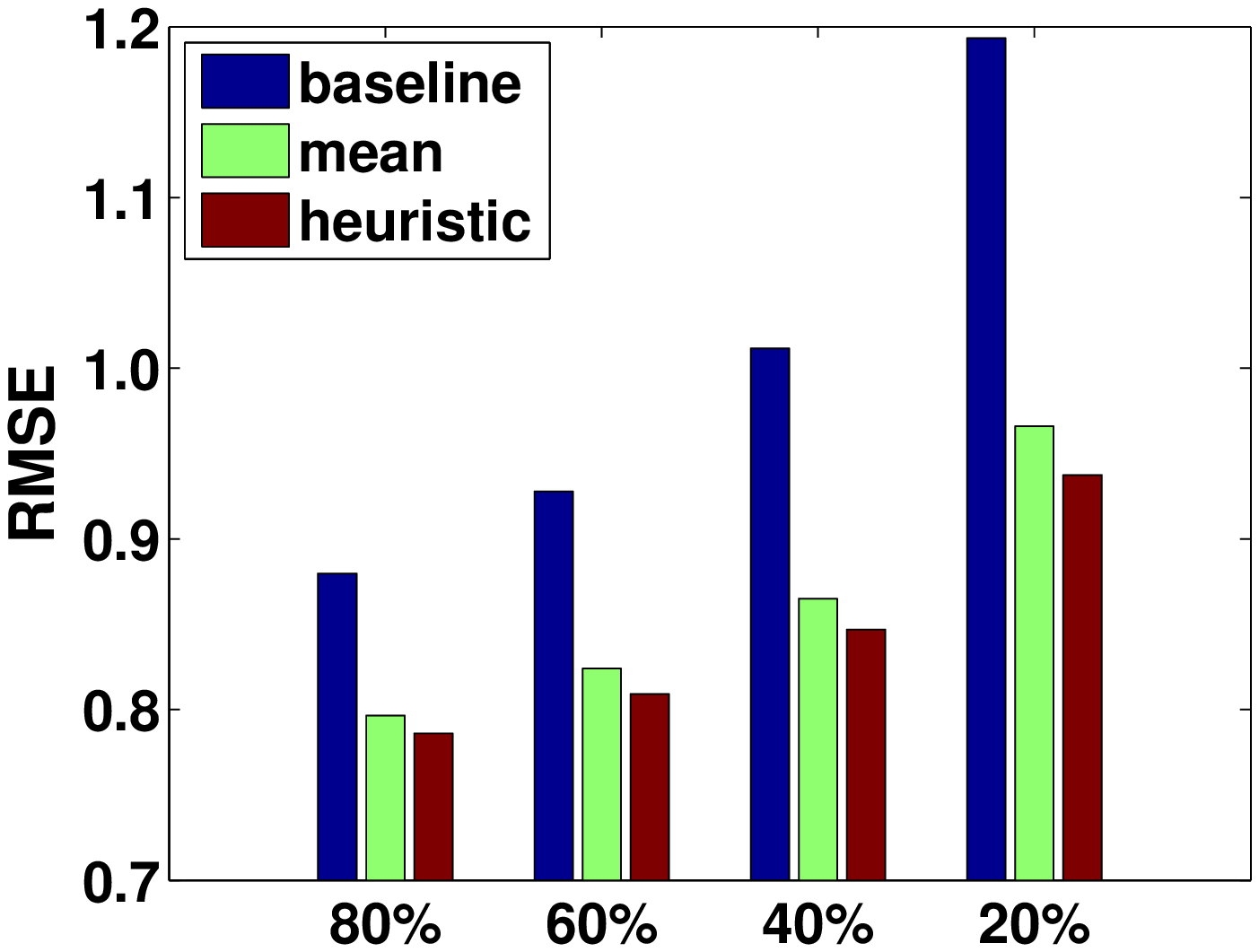}.
\end{minipage}
}\vspace{-10pt}
\caption{\small performance of SimMF on MAE and RMSE with different weights setting methods. } \label{fig:fft}
\end{figure}

\subsection{Parameter Study on $\alpha$ and $\beta$}
Since other parameters have been studied in other matrix factorization methods, here we only do parameter study on $\alpha$ and $\beta$. The parameters $\alpha$ and $\beta$ control how much SimMF fuses the similarity information of users and items. On the one hand, if we only factorize the user-item matrix for recommendation with a very small value of $\alpha$ and $\beta$, SimMF will ignore users' own tastes and items' latent property. On the other hand, if we employ a very large value of $\alpha$ and $\beta$, the similarity information of users and items will dominate the model learning process. Intuitively, we need to set moderate values for $\alpha$ and $\beta$ to balance the rating and similarity information. In this section, we will analyze how the changes of $\alpha$ and $\beta$ can effect the final recommendation accuracy. Specifically, we observe the performance of SimMF-U(a)I(i) with varying $\alpha$ and $\beta$ on Douban Movie dataset.

Figure 5 shows the impacts of $\alpha$ and $\beta$ on MAE and RMSE in SimMF-U(a)I(i) model. We can find that performance of SimMF-U(a)I(i) on MAE and RMSE have very similar trend. Moreover, the value of $\alpha$ and $\beta$ affect recommendation results significantly, which demonstrates that incorporating the similarity information generated by attribute information greatly affects the recommendation accuracy. For very small $\alpha$ and $\beta$, SimMF-U(a)I(i) will degrade to the traditional PMF, which makes its MAE and RSME increase to higher and stable values (i.e., bad performance). For large $\alpha$ and $\beta$, the similarity information of users and items will dominate model learning process, which makes  the MAE and RSME values of SimMF-U(a)I(i) sharply increase. It indicates that the matrix factorization on user-item rating matrix should dominate the learning process, while similarity information is useful supplement to improve performance. In addition, we can observe that, when $\beta$ is around 10 and $\alpha$ is between 10 and 100, SimMF-U(a)I(i) has stable and good performance.
\begin{figure}[htbp]
\centering
\small
\subfigure[MAE]{\label{fig:fft:a}
\begin{minipage}[t]{0.20\textwidth}
  \includegraphics[width=4.5cm]{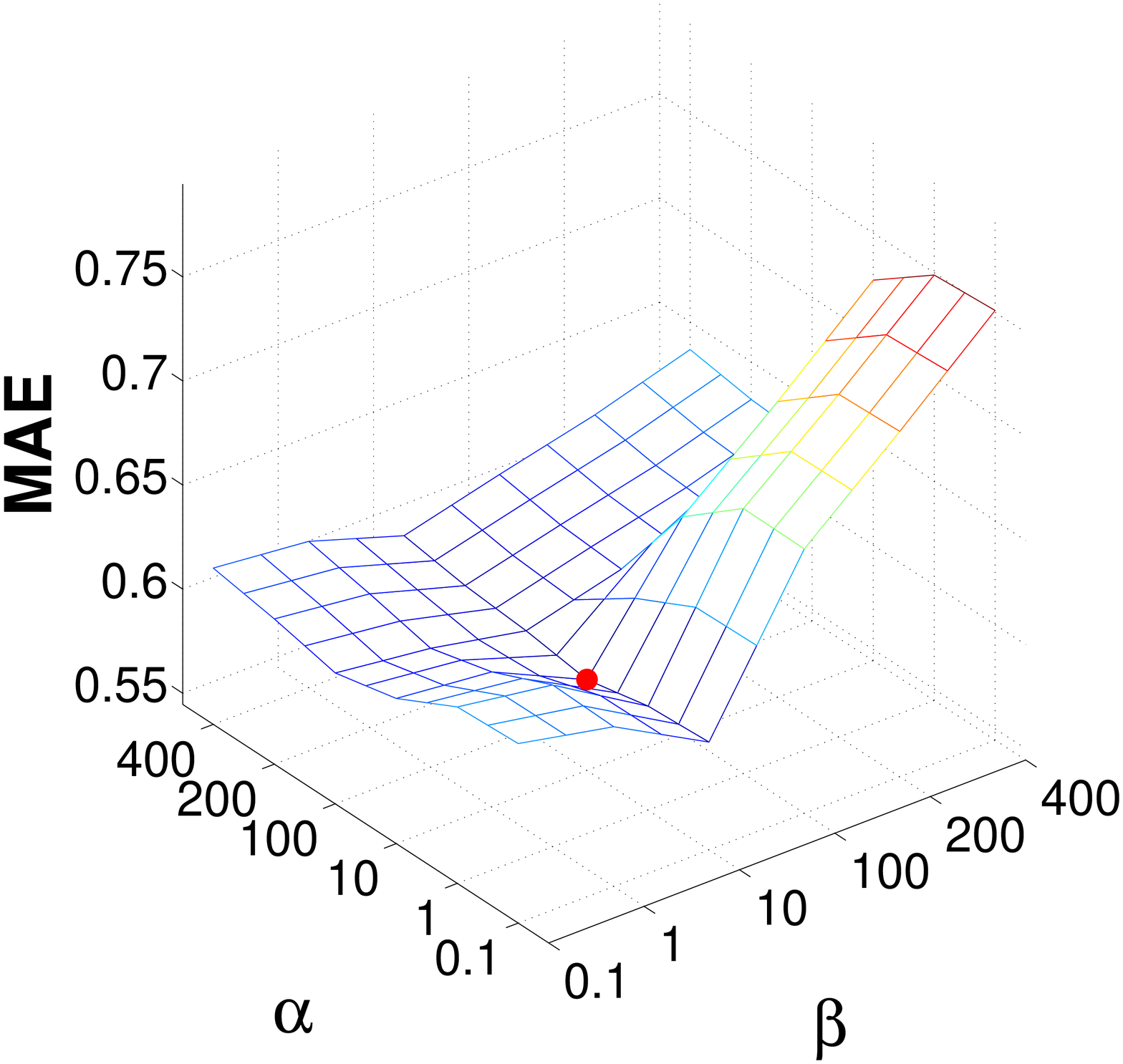}
\end{minipage}%
}
\hspace{0pt}
\subfigure[RMSE]{
\begin{minipage}[t]{0.20\textwidth}
  \includegraphics[width=4.5cm]{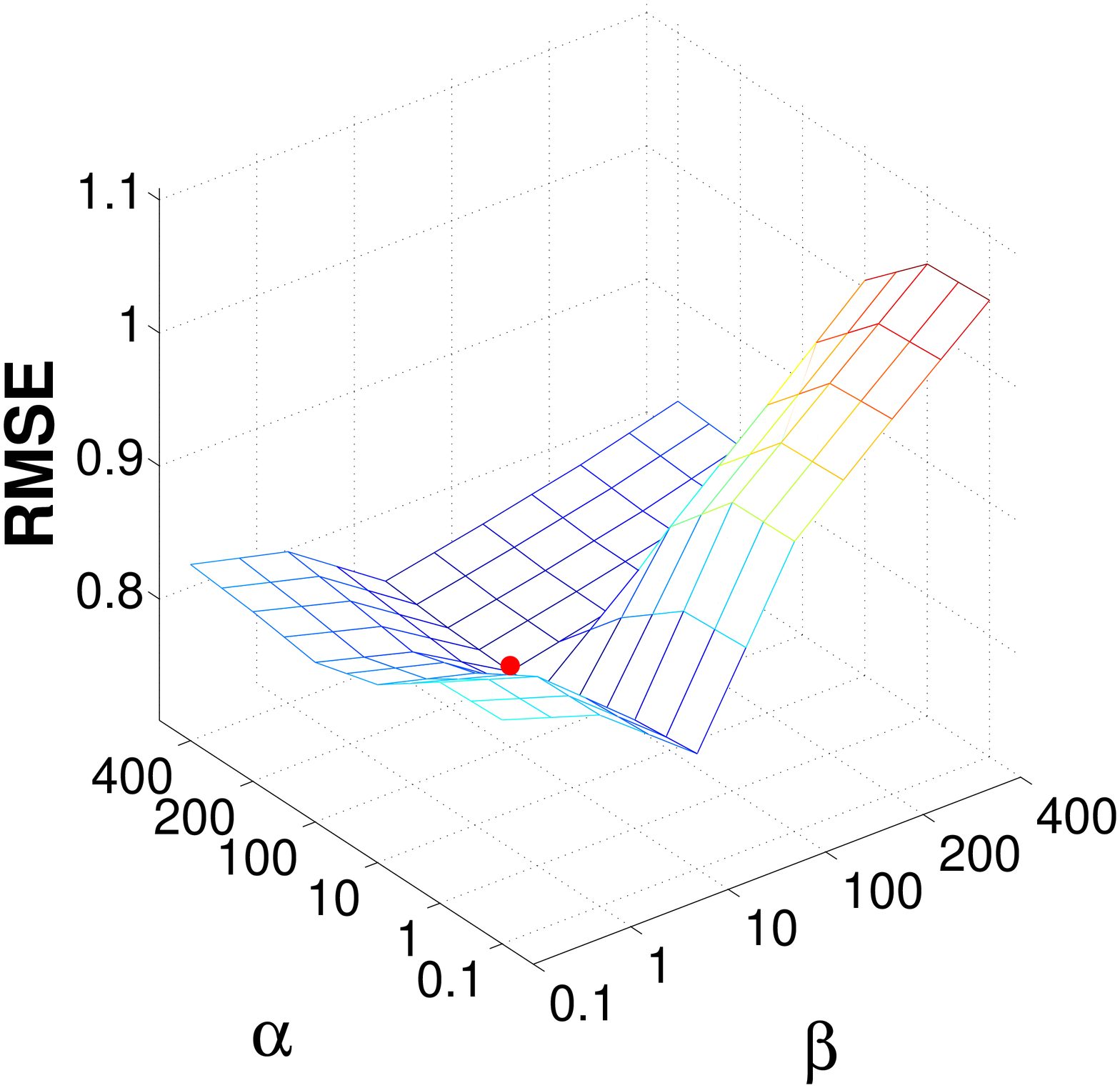}.
\end{minipage}
}\vspace{-10pt}
\caption{\small performance of SimMF on MAE and RMSE with varying $\alpha$ and $\beta$ on Douban Movie dataset. The lower, the better.} \label{fig:fft}
\end{figure}

\section{Conclusion}
In this paper, we organize the objects and relations in recommendation system as a heterogeneous information network, and designed a unified and flexible matrix factorization based dual regularization framework SimMF to effectively integrate different types of information. SimMF employs meta path based similarity measure to evaluate the similarity of objects and flexibly integrate heterogeneous information through adopting the similarity of users and items as regularization on latent factors of user and item. Experiments on real datasets validate the effectiveness of SimMF and some interesting works are needed to explore in the future. It is desirable to design clever weight learning strategy for the combination of similarity matrices to further improve recommendation performance.
\bibliographystyle{abbrv}
\bibliography{References}

\begin{thebibliography}{10}

\bibitem{AT05}
G.~Adomavicius and A.~Tuzhilin.
\newblock Toward the next generation of recommender systems: A survey of the
  state-of-the-art and possible extensions.
\newblock {\em IEEE Transactions on Knowledge and Data Engineering},
  17(6):734--749, 2005.

\bibitem{RIP13}
R.~Bellog\'{i}N, I.~Cantador, and P.~Castells.
\newblock A comparative study of heterogeneous item recommendations in social
  systems.
\newblock {\em Information Sciences}, 221:142--169, 2013.

\bibitem{RFB14}
R.~Burke, F.~Vahedian, and B.~Mobasher.
\newblock Hybrid recommendation in heterogeneous networks.
\newblock In {\em UMAP}, pages 49--60, 2014.

\bibitem{IAD10}
I.~Cantador, A.~Bellogin, and D.~Vallet.
\newblock Content-based recommendation in social tagging systems.
\newblock In {\em RecSys}, pages 237--240, 2010.

\bibitem{FW12}
W.~Feng and J.~Wang.
\newblock Incorporating heterogeneous information for personalized tag
  recommendation in social tagging systems.
\newblock In {\em KDD}, pages 1276--1284, 2012.

\bibitem{JE09}
M.~Jamali and M.~Ester.
\newblock Trustwalker: a random walk model for combining trust-based and
  item-based recommendation.
\newblock In {\em KDD}, pages 397--406, 2009.

\bibitem{JL13}
M.~Jamali and L.~V. Lakshmanan.
\newblock Heteromf: Recommendation in heterogeneous information networks using
  context dependent factor models.
\newblock In {\em WWW}, pages 643--653, 2013.

\bibitem{CJA11}
C.~Jones, J.~Ghosh, and A.~Sharma.
\newblock Learning multiple models for exploiting predictive heterogeneity in
  recommender systems.
\newblock {\em HetRec}, pages 17--24, 2011.

\bibitem{LC10a}
N.~Lao and W.~Cohen.
\newblock Fast query execution for retrieval models based on path constrained
  random walks.
\newblock In {\em KDD}, pages 881--888, 2010.

\bibitem{LHLS15}
H.~Lee and S.-g. Lee.
\newblock Style recommendation for fashion items using heterogeneous
  information network.
\newblock {\em Info Recommender Systems (RecSys), Volume, Page}, 2015.

\bibitem{LPW14}
C.~Luo, W.~Pang, and Z.~Wang.
\newblock Hete-cf: Social-based collaborative filtering recommendation using
  heterogeneous relations.
\newblock In {\em ICDM}, pages 917--922, 2014.

\bibitem{MKL11}
H.~Ma, I.~King, and M.~R. Lyu.
\newblock Learning to recommend with social trust ensemble.
\newblock In {\em SIGIR}, pages 203--210, 2011.

\bibitem{MYLK08}
H.~Ma, H.~Yang, M.~R. Lyu, and I.~King.
\newblock Sorec: Social recommendation using probabilistic matrix
  factorization.
\newblock In {\em CIKM}, pages 931--940, 2008.

\bibitem{MZLLK11}
H.~Ma, D.~Zhou, C.~Liu, M.~R. Lyu, and I.~King.
\newblock Recommender systems with social regularization.
\newblock In {\em WSDM}, pages 287--296, 2011.

\bibitem{MZLK11}
H.~Ma, T.~Zhou, M.~Lyu, and I.~King.
\newblock Improving recommender systems by incorporating social contextual
  information.
\newblock {\em ACM Transactions on Information Systems}, 29(2):9, 2011.

\bibitem{SM95}
U.~Shardanand and P.~Maes.
\newblock Social information filtering: Algorithms for automating word of
  mouth.
\newblock In {\em Conf. Human Factors in Computing Systems}, 1995.

\bibitem{SKHYW14}
C.~Shi, X.~Kong, Y.~Huang, P.~S. Yu, and B.~Wu.
\newblock Hetesim: A general framework for relevance measure in heterogeneous
  networks.
\newblock {\em IEEE Transactions on Knowledge and Data Engineering}, pages
  2479--2492, 2014.

\bibitem{SZLYYW15}
C.~Shi, Z.~Zhang, P.~Luo, P.~S. Yu, Y.~Yue, and B.~Wu.
\newblock Semantic path based personalized recommendation on weighted
  heterogeneous information networks.
\newblock In {\em Proceedings of the 24th ACM International on Conference on
  Information and Knowledge Management}, pages 453--462. ACM, 2015.

\bibitem{SJ03}
N.~Srebro and T.~Jaakkola.
\newblock Weighted low-rank approximations.
\newblock In {\em ICML}, pages 720--727, 2003.

\bibitem{SHYYW11}
Y.~Sun, J.~Han, X.~Yan, P.~Yu, and T.~Wu.
\newblock Pathsim: Meta path-based top-k similarity search in heterogeneous
  information networks.
\newblock In {\em VLDB}, pages 992--1003, 2011.

\bibitem{Sun4}
Y.~Sun, B.~Norick, J.~Han, X.~Yan, P.~S. Yu, and X.~Yu.
\newblock Integrating meta-path selection with user guided object clustering in
  heterogeneous information networks.
\newblock In {\em KDD}, pages 1348--1356, 2012.

\bibitem{FV14}
F.~Vahedian.
\newblock Weighted hybrid recommendation for heterogeneous network.
\newblock In {\em RecSys}, pages 429--432, 2014.

\bibitem{YSL12}
X.~Yang, H.~Steck, and Y.~Liu.
\newblock Circle-based recommendation in online social networks.
\newblock In {\em KDD}, pages 1267--1275, 2012.

\bibitem{YMHH14}
X.~Yu, H.~Ma, B.-J.~P. Hsu, and J.~Han.
\newblock On building entity recommender systems using user click log and
  freebase knowledge.
\newblock In {\em WSDM}, pages 263--272, 2014.

\bibitem{YRGSH13}
X.~Yu, X.~Ren, Q.~Gu, Y.~Sun, and J.~Han.
\newblock Collaborative filtering with entity similarity regularization in
  heterogeneous information networks.
\newblock In {\em IJCAI-HINA workshop}, 2013.

\bibitem{YRSGSKNH14}
X.~Yu, X.~Ren, Y.~Sun, Q.~Gu, B.~Sturt, U.~Khandelwal, B.~Norick, and J.~Han.
\newblock Personalized entity recommendation: a heterogeneous information
  network approach.
\newblock In {\em WSDM}, pages 283--292, 2014.

\bibitem{YRSSKGNH13}
X.~Yu, X.~Ren, Y.~Sun, B.~Sturt, U.~Khandelwal, Q.~Gu, B.~Norick, and J.~Han.
\newblock Recommendation in heterogeneous information networks with implicit
  user feedback.
\newblock In {\em RecSys}, pages 347--350, 2013.

\bibitem{ZTL08}
J.~Zhang, J.~Tang, B.~Liang, Z.~Yang, S.~Wang, J.~Zuo, and J.~Li.
\newblock Recommendation over a heterogeneous social network.
\newblock In {\em Web-Age Information Management, 2008. WAIM'08. The Ninth
  International Conference on}, pages 309--316. IEEE, 2008.

\end{thebibliography}

\end{document}